\documentclass[a4paper,12pt]{article}
\usepackage{jcappub} % for details on the use of the package, please
\usepackage{cite}
\usepackage{graphicx}
\usepackage{enumerate}
\usepackage{hyperref}
\usepackage{cancel}
\usepackage{color}
\usepackage{graphicx}
\usepackage{amsmath}
\usepackage{amsfonts}
\usepackage{amssymb}
\usepackage{rotating}
\usepackage{booktabs}
\usepackage{xcolor}
\usepackage{soul}

\def\comment#1{}

\title{\boldmath 
Holographic massive plasma state in Friedman Universe:
cosmological fine-tuning and coincidence problems
}

\author{She-Sheng Xue}

\affiliation{ICRANet Piazzale della Repubblica, 10 -65122, Pescara, Italy
\\ Physics Department, Sapienza University of Rome, 
Rome, Italy\\INFN, Sezione di Perugia, %Via A. Pascoli, I-06123, 
Perugia, Italy
\\ICTP-AP, University of Chinese Academy of Sciences, Beijing, China
}

\emailAdd{xue@icra.it, she-sheng.xue@cern.ch} 

\abstract{Massive particle and antiparticle pair production and 
oscillation on the horizon form a holographic and massive pair plasma state in the Friedman Universe.  Via this state, the Einstein 
cosmology term (dark energy) interacts with matter and radiation and is time-varying $\tilde\Lambda$ in the Universe's evolution. 
It is determined by a close set of ordinary differential equations for dark energy, matter, and radiation energy densities. The solutions are unique, provided the initial conditions given by observations. 
In inflation and reheating, dark energy density decreases from the inflation scale, converting to matter and radiation energy densities. 
In standard cosmology, matter and radiation energy densities convert to dark energy density, reaching the present Universe. By comparing with $\Lambda$CDM, quintessence and dark energy interacting models, we show that 
these results can be the possible solutions for cosmological fine-tuning and coincidence problems. 
}

\begin{document}
\maketitle
\flushbottom

%%%%%%%%%%%%%%%%%%%%%%%%%%%%%%%%%%%%%%%%%%%%%%%%%%%%%%%%%%%%%%%%%%%%%%%%%%%%%%%%%%%%%%%%%%%%%%%%%%%%%%%%%%%%%%%%%%%%%%%%%%%%
%%%%%%%%%%%%%%%%%%%%%%%%%%%%%%%%%%%%%%%%%%%%%%%%%%%%%%%%%%%%%%%%%%%%%%%%%%%%%%%%%%%%%%%%%%%%%%%%%%%%%%%%%%%%%%%%%%%%%%%%%%%%
%%%%%%%%%%%%%%%%%%%%%%%%%%%%%%%%%%%%%%%%%%%%%%%%%%%%%%%%%%%%%%%%%%%%%%%%%%%%%%%%%%%%%%%%%%%%%%%%%%%%%%%%%%%%%%%%%%%%%%%%%%%%
%

\section{\bf Introduction}\label{int}
%\noindent{\bf Introduction~}\label{int}
%\hskip0.1cm

Einstein's equation describes the Universe's evolution in terms of Newton's constant $G$, Ricci scalar $R$,  cosmological constant $\Lambda$ and matter and radiation energy densities. Einstein introduced the $\Lambda$ to counterbalance attractive gravity and achieve a static universe. It was abandoned after the confirmation of Universe expansion. It has been revived since the discovery of the Universe's acceleration, which implies that the $\Lambda$ may have a positive value. 
Owing to its mysterious origin and nature, the $\Lambda$ component is dubbed as ``dark energy''. As a standard model of the Big Bang cosmology, the simplest $\Lambda$CDM (Lambda cold dark matter) model provides a reasonably good account of the observed properties of the cosmos and accelerating Universe. 

However, whatever the dark energy form and nature are, two cosmological problems arise in $\Lambda$CDM. The cosmic fine-tuning problem \cite{Weinberg1989}: how to explain the present dark energy density $\sim 10^{-47} {\rm GeV}^4$ is about $10^{-123}$ smaller than the energy density at the Planck scale $M_{\rm pl}=G^{-1/2}\approx 1.2\times 10^{19} $GeV. It is a unique basic scale entering Einstein's equation. The cosmic coincidence problem \cite{Zlatev1999, Huey2006, Velten2014}: how to explain the coincidence of dark energy and matter densities at the present epoch after a long cosmic evolution. Observe that throughout the cosmic evolution after the Big Bang (reheating), dark energy density $\rho_{_\Lambda}=\Lambda^2/(8\pi G)$
does not change. While radiation and matter energy densities fall in powers of the scalar factor $a$ and change many orders of magnitudes. In the recent epoch, their abundances $\Omega_\Lambda\approx 0.7$ and $\Omega_M\approx 0.3$ not only coincide in order of magnitude but also are just the correct values for forming structure, galaxy, and astrophysical objects. It is just the world where human beings create and live. There are three possibilities. (i) We are subject to the Anthropic Principle and happen to live with a peculiar $\Lambda$ value and in a special epoch after the long history of Big Bang cosmology, (ii) Nature fine-tunes in many orders of magnitude the $\Lambda$ value and the ratio of dark energy and radiation densities at the beginning of Big Ban.
(iii) Dark energy is time-varying due to dynamically interacting with matter and radiation.

The cosmic fine-tuning and coincidence problems have been studied in many cosmological model extensions to  $\Lambda$CDM, for example, quintessence  \cite{Caldwell1998, Steinhardt2005, Amendola2000, Amendola2015}, interacting dark energy model \cite{ Boehmer2008, Valiviita2008, Campo2009, Wang2016, Bolotin2014, DiValentino2020, DiValentino2020a, Pan2020, Pan2020a, Lucca2020, Lucca2021}, phenomenological model \cite{Dalal2001}.  People have not yet found consistent solutions to these two problems in the standard cosmology after Big Ban. 
In addition, these beyond $\Lambda$CDM scenarios have not yet given an overall consistent description of inflation, reheating, and standard cosmology epochs.
In the recently proposed $\tilde\Lambda$CDM
model of time-varying $\tilde\Lambda$ due to dark energy and matter interactions, we describe inflation epoch \cite{Xue2023} and reheating epoch \cite{Xue2023a}.  Following these studies, in this article, 
we try to find and explain 
a self-consistent dynamical solution for the cosmic fine-tuning and coincidence problems in the standard cosmology epoch.

First, we briefly review the $\tilde\Lambda$CDM model and its applications to the inflation and reheating epoch.
\comment{, baryogenesis and magnetogenesis phenomena.} 
Then,
we present the Friedman equations for the Hubble function $H$ and time-varying dark energy $\rho_{_\Lambda}$, the cosmic rate equations for matter $\rho_{_M}$ and radiation $\rho_{_R}$ densities that interact with dark energy density $\rho_{_\Lambda}$. These equations form a close set of ordinary differential equations. The solution is unique, provided initial conditions are given by observations.  
Numerically integrating these equations, we find the $\tilde\Lambda$CDM dynamical solution: (i) in inflation and reheating, dark energy converts to matter and radiation energies and vanishes at the end of reheating; (ii) in standard cosmology, instead,  
matter and radiation energies convert to dark energy. The conversion rate is proportional to $1/H$, namely dark energy and matter interaction decreases as redshift $z$ increases. These dynamical features yield a possible solution for the cosmic coincidence problem in $\Lambda$CDM. The conversion rate $\propto 1/H$ is consistent with the late-time interaction
in dark matter converting dark energy, obtained by data analysis in different $z$ value bins\cite{Salvatelli2014, Gariazzo2022}.

We compare and contrast $\tilde\Lambda$CDM with $\Lambda$CDM and advocate an approximate model for phenomenological studies and data analysis. 
Finally, we discuss in the $\tilde\Lambda$CDM scenario 
the geometric and dynamic natures of $\tilde\Lambda$ dark energy as a gravitational ground state, asymptotically safe Einstein theory for the early and present Universe, and $\tilde\Lambda$ dark energy solution for the cosmic fine-tuning problem.
%In this article, $G=M^{-2}_{\rm pl}$ is the the Newton constant, $M_{\rm pl}$ is the Planck scale and reduced Planck scale $m_{\rm pl}\equiv (8\pi)^{-1/2} M_{\rm pl}=2.43\times 10^{18} $GeV.

\comment{
rather than an inverse process in other dark energy and matter interacting models \cite{Campo2009}. 
Many theoretical ideas have been motivated for cosmology,
and advocated examining the $H_0$ tension 
recently observed 
Refs.~\cite{DiValentino2017a,DiValentino2021b,Verde2019,DiValentino2021c,Freedman2017,Riess2019,Camarena2018,Salvatelli2013,Costa2014,Li2013,Yang2017,DiValentino2021,Yang2021,DiValentino2017,Zhao2017,Martinelli2019,Alestas2020,DiValentino2021a,Efstathiou2020,Yang2021a,Huang2016}.
references:
summary of interacting dark energy  https://arxiv.org/pdf/2302.11949.pdf
Gregory: https://arxiv.org/pdf/2208.04596.pdf
https://arxiv.org/abs/1203.4197
https://arxiv.org/pdf/2107.08916.pdf
}

\section{\bf Cosmological $\tilde\Lambda$CDM model }\label{rlcdm}
 
In such $\tilde\Lambda$CDM scenario, we have recently studied singularity-free and large-scale anomaly issues, the spectral index and tensor-to-scalar ratio relation in the inflation epoch \cite{Xue2023}, and 
calculated reheating energy and entropy \cite{Xue2023a}. The results are consistent
with observations. We briefly recall three main features of the $\tilde\Lambda$CDM scenario.

\subsection{Time-varying cosmological $\tilde\Lambda$ term}
First, a time-varying cosmological $\tilde\Lambda$ term represents interacting dark energy with matter and radiation. 
The Friedman equations for a flat Universe of horizon $H$ are \cite{Xue2015}
\begin{eqnarray}
H^2 &=& \frac{8\pi G}{3}(\rho_{_M}+\rho_{_R}+\rho_{_\Lambda}),\label{friedman0}\\
\dot H &=&-\frac{8\pi G}{2}(\rho_{_M}+\rho_{_R}+\rho_{_\Lambda} + p_{_M}+p_{_R}+p_{_\Lambda}).
\label{friedman}
\end{eqnarray}
Equations of States $p_{_{M,R,\Lambda}}= \omega_{_{M,R,\Lambda}} \rho_{_{M,R,\Lambda}}$, $\omega_{_{M}}= 0$ for massive particles and $\omega_{_{R}}= 1/3$ for massless radiation.
The second equation of (\ref{friedman}) is the generalized 
conservation law (Bianchi identity) for including time-varying cosmological term
$\rho_{_\Lambda}(t)\equiv \tilde\Lambda(t)/(8\pi G)$ and $\omega_{_\Lambda}=-1$. When $\Lambda$ is constant in time, particles are stable, and dark energy does not interact with matter and radiation, Equation (\ref{friedman}) reduces to the usual equations 
$\dot \rho_{_\Lambda}=0$, $\dot \rho_{_M} + 3H\rho_{_M}=0$ and  $\dot \rho_{_R} + 4H\rho_{_R}=0$, whose solutions in terms of scale factor $a$ are $(1/a)^{3(1+\omega_{_{\Lambda,M,R}})}$, respectively for constant dark energy density, massive particle number and massless particle number
(entropy) conservation. When interacting $\tilde\Lambda$ dark energy is time-varying, Eqs.~(\ref{friedman0},\ref{friedman}) lead to $\dot \rho_{_\Lambda}+\dot \rho_{_R}+\dot \rho_{_M}=-H(3\rho_{_R}+4\rho_{_R})$, whose solutions $\rho_{_{\Lambda,M,R}}$ differ from the usual one. For weak $\tilde\Lambda$ interacting with matter and radiation, we parameterize the solutions
$\rho_{_{\Lambda,M,R}}\propto (1/a)^{3(1+\omega^{\rm eff}_{_{\Lambda,M,R}})}$ with effective equation of state $\omega^{\rm eff}_{_{\Lambda,M,R}}= \omega_{_{\Lambda,M,R}} + \delta_{_{\Lambda,M,R}}$ and $|\delta_{_{\Lambda,M,R,}}|\ll 1$ \cite{Begue2019}, see also Sec.~\ref{appL}. 
The detailed discussions are in Secs.~7 and 9 of Ref.~\cite{Xue2015}.    

\subsection{Massive pair production and oscillation}

Second, the spontaneous gravitational creation of massive particle and anti-particle pairs in the Friedman Universe has been intensively studied, see examples \cite{Parker1973, Starobinsky1982, Ford1987, 
%Kolb1996,
Chung2019,
%Ema2018,Li2019,Xue2019,
Xue2020}. The massive pairs' production \cite{Parker1973} and oscillation \cite{Xue2023, Xue2023a}  establish a condensate ground state $|{\mathcal N}_{\rm pair}\rangle$ of the large number (${\mathcal N}_{\rm pair}\gg 1$) and massive ($M\gg H$) pairs of particles and anti-particles. They attribute to the microscopic
fast-component $H_{\rm fast}$ in the Hubble function $H=H_{\rm fast}+H_{\rm slow}$. The fast component $H_{\rm fast}$ oscillates coherently with the pairs' oscillation, which relates to pairs' production and annihilation from/into the vacuum at the microscopic time scale $1/M$. 
We present in Appendix \ref{qppos} the local and fast-oscillating $H_{\rm fast}$ dynamics. 
It is consistent with recent studies of vacuum fluctuation and ``microcyclic universes'' at small scales, shown by local scale factor oscillation 
in Figure 1 of Refs.~\cite{Wang2020, Wang2020a}. 
The macroscopic slow component  $H_{\rm slow}\gg H_{\rm fast}$, and $H_{\rm slow}\approx H$ obeys the Friedman equation 
(\ref{friedman}) at the ``macroscopic'' time scale $1/H$.
These microscopic and macroscopic 
processes couple each other.
However, one cannot even numerically integrate their differential equations due to the vast difference between the scale $1/M$ and $1/H$. Therefore, at the macroscopic time scale $1/H$, averaging over microscopic states, we model the condensate ground state $|{\mathcal N}_{\rm pair}\rangle$ as an ``equilibrium'' or ``equipartition'' state of the microscopic fast component $H_{\rm fast}$ and pairs' oscillation. It is the method that we use to study the back-reactions of microscopic fast component $H_{\rm fast}$ and pairs' oscillation on macroscopic densities $\rho_{_{\Lambda, M, R}}$ in the Friedman equation (\ref{friedman}). Detailed discussions are in Secs.~2-3 of Ref.~\cite{Xue2023a}.

\subsection{Holographic and massive
pair plasma state}

Third, we assume the aforementioned ``equilibrium'' state $|{\mathcal N}_{\rm pair}\rangle$ is a holographic and massive pair plasma state containing a large number 
of massive particle and antiparticle pairs. We effectively describe such a plasma state as a perfect fluid state of effective number $n^H_{_M}$ and 
energy $\rho^H_{_M}$ densities of stable and unstable massive pairs,
\begin{eqnarray}
%\langle\rho^{\rm fast}_{_M}+ p^{\rm fast}_{_M}\rangle\Rightarrow 
\rho^H_{_M} \equiv  2\chi  m^2 H^2,\quad n^H_{_M} \equiv   \chi  m H^2. %;\quad m^2 \equiv \sum_fg_d^fM^2_f,
\label{apdenm}
\end{eqnarray}
The equation of state and pressure are $p^H_{_M}=\omega^H_{_M}
\rho^H_{_M}$. The lower limit $\omega^H_{_M}\approx 0$ for $m\gg H$,
and the upper limit  $\omega^H_{_M}\lesssim 1/3$ for $m\gtrsim H$. The $m\propto {\mathcal N}_{\rm pair}M$ is the effective mass parameter, representing total mass and number of pairs in the massive pair plasma state (\ref{apdenm}). Such a state is a 
holographic layer near the horizon because the average of local pairs' and $H_{\rm fast}$ oscillations at the scale $1/M$ should vanish inside the horizon for global homogeneity up to the horizon, as explained in Sec.~4 of Ref.~\cite{Xue2023a}. The width parameter $\chi$ characterizes the layer radial width $\lambda_m = (\chi m)^{-1}$. We adopt \footnote{In Refs.~\cite{Xue2019, Xue2020}, we adopt the different renormalization prescription 
at high energies $M\gg H$ from the usual prescription (subtraction) at low energies $M\ll H$. We have consistently obtained the mean density $n^H_{_M} \approx \chi m H^2$ (\ref{apdenm}) 
and $\chi\approx 1.85\times 10^{-3}$ by studying massive fermion pair productions in an exact De Sitter spacetime of constant $H$ and scaling factor $a(t)=e^{iHt}$. This result implies $\chi\sim {\mathcal O}(10^{-3})$.} $\chi=10^{-3}$ and treat $m$ as a free parameter.
%The conditions $\lambda_m < H^{-1}$ and $\lambda_m \gg 1/m$ gives the constraint $1\gg \chi > H/m$ 
At a given horizon $H$, the ``macroscopic'' condensation state $\rho^H_{_M}$ (\ref{apdenm}) effectively represents the average overall ``microscopic'' states of pair production, annihilation and oscillations at the time scale $1/M$. Since produced pairs' mass $M$ and number ${\mathcal N}_{\rm pair}$ cannot be constant in time, we assume the effective mass parameter $m$ weakly depends on the horizon $H$. Its effective value $m_{\rm eff}$ for each evolution epoch must be fixed by observations.
The detailed discussions are in Secs.~3-4 of Ref.~\cite{Xue2023a}. 

\subsection{Cosmic rate equations for matter and radiation densities}

Fourth, we describe how the massive pair plasma energy density $\rho^H_{_M}$ interacts with matter and radiation energy density $\rho_{_{M, R}}$, which vary in time scale $1/H$. While the $\rho^H_{_M}$ variation time scale $\tau_{_M}$ differs from $1/H$ and can be estimated as follows.
From the pair number density $n^H_{_M}$ (\ref{apdenm}),
the total number of particles produced inside the Hubble sphere $N\approx n^H_{_M}H^{-3}/2$ and mean pair production rate w.r.t.~macroscopic time variation $dt$ are approximately,
\begin{eqnarray}
\Gamma_M &=& \frac{dN}{2\pi dt}\approx \frac{\chi m}{4\pi} \epsilon, \quad \tau^{-1}_{_M}=\Gamma_M. \label{prate} 
\end{eqnarray}
The Universe evolution $\epsilon$-rate is usually defined as,
\begin{eqnarray}
\epsilon &\equiv& -\frac{\dot H}{H^2} =\frac{3}{2}\frac{(1+\omega_{_M})\rho_{_M}+ (1+\omega_{_R})\rho_{_R}+(1+\omega_{_\Lambda})\rho_{_\Lambda}}{\rho_{_\Lambda}+\rho_{_M}+\rho_{_R}},
\label{dde0}
\end{eqnarray}
where the second equality comes from the Friedman equations (\ref{friedman}).
The asymptotic values $\epsilon \ll  1$, $\epsilon \approx  2$, and $\epsilon \approx  3/2$ correspond to dark energy, radiation, and matter domination, respectively. 

The massive pair plasma state density $\rho^H_{_M}$ associated with the horizon contributes to the matter/radiation density $\rho_{_{M, R}}$ in Friedman equations (\ref{friedman0}) and (\ref{friedman}). In turn, the $\rho_{_{M, R}}$ variation affects the $\rho^H_{_M}$ via the horizon $H$. 
It implies the back-and-forth interaction between the massive pair plasma state and the matter/radiation state
during the Universe's evolution. Moreover, the massive pair plasma state $\rho^H_{_M}$ (\ref{apdenm}) has a microscopic ``relaxation'' time scale $\tau_{_M}=\Gamma_M^{-1}$ (\ref{prate}), differing from the macroscopic one $\tau_{_H}=H^{-1}$ of the matter/radiation state 
$\rho_{_{M, R}}$ in Friedman equations (\ref{friedman0}) and (\ref{friedman}), i.e., $\tau_{_H}\gg \tau_{_M}$. Therefore, we cannot simply add $\rho^H_{_M}$ into $\rho_{_{M, R}}$ in Friedman equations. 

By analogy with the rate equation for a microscopic back-and-forth process, e.g., $e^+e^-\Leftrightarrow\gamma\gamma$, in macroscopic expansion (see part of Eq.~(5.4) in Ref.~\cite{Xue2023a}), we propose the back-and-forth interaction between the densities $\rho^H_{_M}$ and $\rho_{_{M,R}}$ follows the cosmic 
rate equations of Boltzmann type,
\begin{eqnarray}
\dot\rho_{_M}+ 3(1+\omega_{_M}) H\rho_{_M} &=& \Gamma_M(\rho_{_M}^H - \rho_{_M}-\rho_{_{R}}) - \Gamma_M^{^{\rm de}}\rho^{\rm de}_{_M},
\label{rateeqd}\\
\dot\rho_{_{R}}+ 3(1+\omega_{_{R}}) H\rho_{_{R}} &=& \Gamma_M(\rho_{_M}^H - \rho_{_{M}}-\rho_{_{R}})+\Gamma_M^{^{\rm de}}\rho^{\rm de}_{_{M}}.
\label{rateeqr}
\end{eqnarray}    
The term $3(1+\omega_{_{M,R}}) H\rho_{_{M,R}}$ of the time scale $[3(1+\omega_{_{M,R}}) H]^{-1}$ represents the space-time expanding effect on the density $\rho_{_{M,R}}$. The detailed balance term 
$\Gamma_M(\rho_{_M}^H - \rho_{_{M}}- \rho_{_{R}})$ indicates how densities $\rho_{_M}^H$ and $\rho_{_{M,R}}$ of different time scales couple together in back-and-forth interaction. 
$\Gamma_M \rho_{_M}^H$ is the source term, indicating 
$\rho_{_M}^H$ contribution to increasing $\rho_{_{M,R}}$. 
$\Gamma_M(\rho_{_{M}}+\rho_{_{R}})$ is the depletion term, indicating back reaction reducing $\rho_{_{M,R}}$. 
The ratio $\Gamma_M/H> 1$ indicates the coupled case, and  $\Gamma_M/H < 1$ indicates the decoupled case.  
There are unstable massive pairs of density $\rho^{\rm de}_{_M}$ inside the massive pair plasma state (\ref{apdenm}). The term $\Gamma^{^{\rm de}}_M\rho^{\rm de}_{_M}$ represents unstable massive pairs decay to light particles, such as quarks and leptons, gauge bosons in SM and other light sterile particles. The detailed discussions are in Secs.~4-5 of Ref.~\cite{Xue2023a}.

\subsection{Preliminary applications to inflation and reheating}
The main aspects of the $\tilde\Lambda$CDM scenario are (a) the dark energy and matter interacting Friedman Equations (\ref{friedman0},\ref{friedman}); (b) massive particle and antiparticle pairs' production and oscillation; (c) a holographic massive pair plasma state (\ref{apdenm}) and its variation rate (\ref{prate}); (d) cosmic rate equations (\ref{rateeqd}) and (\ref{rateeqr}). 
They form a close set of first-order ordinary differential equations for the densities
$\rho_{_{M}}, \rho_{_{R}}, \rho_{_{\Lambda}}$ and Hubble function $H$. The solutions are completely determined, provided initial or transition conditions and effective mass parameter $m_{\rm eff}$ are fixed by observations.

In Ref.~\cite{Xue2023}, we study the inflation when the dominant dark energy $\rho_{_\Lambda}$ drives inflation and produces massive pairs' plasma $\rho^H_{_M}$, that 
contributes to the matter $\rho_{_{M}}$ and 
slows down inflation ($\rho_{_\Lambda}\gg \rho^H_{_M}\approx \rho_{_{M}}$). 
Neglecting Eqs.~(\ref{rateeqd},\ref{rateeqr}) for the decoupled case $\Gamma_M/H<1$, we 
approximately use Eq.~(\ref{friedman}) and $\rho_{_M}\approx \rho^H_{_{M}}$ to obtain an analytical solution $
H_{\rm end}=H_*\exp -(\epsilon^*\, N_{\rm end}).
$
The $e$-folding numbers $N_{\rm end} \approx (50~60)$ from the CMB pivot scale $H_*$ to the inflation end $H_{\rm end}\approx \Gamma_M\approx(0.42,0.35)H_*$. In the inflation epoch, we fix the effective value of mass parameter
$m_{\rm eff}=m_*$ by $\epsilon^* =\chi (m_*/m_{\rm pl})^2=(1-n_s)/2$ \footnote{The reduced Planck mass $m_{\rm pl}\equiv (8\pi)^{-1/2} M_{\rm pl}=2.43\times 10^{18} $GeV.}.
The obtained relation of spectral index $n_s$ and tensor-to-scalar ratio $r$ agrees with
recent CMB observations.
We discuss the singularity-free pre-inflation, the CMB large-scale
anomaly, and dark-matter density perturbations imprinting on power spectra.

In Ref.~\cite{Xue2023a}, we study the
reheating when the dark energy $\rho_{_\Lambda}$ decreases, $\rho^H_{_M}$ and $\rho_{_M}$ increase. The competition between the Hubble function $H$, 
the massive pair plasma variation rate $\Gamma_M$ (\ref{prate}) and massive pairs' decay rate $\Gamma_M^{^{\rm de}}$ play an important role in cosmic rate equations (\ref{rateeqd}) and (\ref{rateeqr}). First, it appears the ${\mathcal M}$-episode of massive pair domination when $\Gamma_M > H>\Gamma_M^{^{\rm de}}$. Then it proceeds to the ${\mathcal R}$-episode of radiation domination when $\Gamma_M^{^{\rm de}}> H>\Gamma_M$. The cosmic rate equation (\ref{rateeqr}) becomes a reheating equation for $\Gamma_M^{^{\rm de}}>\Gamma_M$. Unstable pairs decay to light particles, and the radiation energy $\rho_{_R}$ increases, leading to reheating.
%at the scale $H_{\rm RH}$ and temperature $T_{\rm RH}\approx H_{\rm RH}$. 
In the reheating epoch, we fix the effective value of mass parameter
$m_{\rm eff}=\hat m\gtrsim 20 ~m_{\rm pl}$ and obtained results %calculate the ratio $H_{\rm RH}/H_{\rm end}\sim 10^{-4}$ 
agree with observations. Stable massive particles remain as cold dark matter particles \footnote{There, the strongly coupled case $\Gamma_M/H \gg 1$ is assumed in the preliminary study of cold dark matter abundance $\Omega_M$ evolution. 
We realise it should be the weakly coupled case $\Gamma_M/H < 1$ after studying the $\Omega_M$ evolution in this article.}. The detailed discussions are in Secs.~7.2-7.3 of Ref.~\cite{Xue2023a}.  

\comment{In Ref.~\cite{Xue2020b}, we discuss particle and
antiparticle densities perturbations in the massive pair plasma that form the acoustic waves of particle-antiparticle symmetric and asymmetric densities in the ${\mathcal M}$-episode. Comparing
their wavelength with the horizon size, we show the asymmetry of massive particles and antiparticles due to the superhorizon crossing. It leads to baryogenesis and magnetogenesis, and the obtained baryon number-to-entropy ratio and primordial magnetic field upper and lower limits agree with observations.
Moreover, we study the physically interested perturbation modes that represent dark-matter acoustic waves. These modes exited from the horizon
and returned to the horizon after the recombination. Thus, they possibly imprint on
the matter power spectrum at large length scales. They have physical influences on the
formation of large-scale structures and galaxies.
}

\section{$\tilde\Lambda$CDM equations of dark energy and matter interaction}\label{coin}

At the reheating end, the radiation energy density $\rho_{_R}$ is dominant, stable cold dark matter energy density $\rho_{_M}\ll \rho_{_R}$ and the dark energy density nearly vanishes $\rho_{_\Lambda}\approx 0$, namely $\rho_{_R}\gg \rho_{_M}\gg \rho_{_\Lambda}\approx 0$. These are the initial conditions starting the standard cosmology. It then evolves to matter-dominated, dark energy-dominated epochs.
We study in this article the $\tilde\Lambda$CDM equation and solution after reheating, focusing on the problem of cosmological coincidence between dark energy and matter.
%A large variation of the Hubble function from the reheating end $H_{\rm RH}$ to the present time $H_0$. 

\subsection{Dark energy interaction with matter and radiation}

To explicitly show dark energy and matter interaction, we recast the Friedman equations (\ref{friedman0},\ref{friedman}), cosmic rate equations (\ref{rateeqd}) and (\ref{rateeqr}) as
\begin{eqnarray}
\dot\rho_{_\Lambda}+3(1+\omega_{_\Lambda}) H\rho_{_\Lambda}&=& - 2 \Gamma_M \left(\rho^H_{_M} - \rho_{_{M}}-\rho_{_{R}}\right),
\label{rhoL}\\
\dot\rho_{_{M}} + 3(1+\omega_{_M}) H\rho_{_{M}}&=& +\Gamma_M \left(\rho^H_{_M} - \rho_{_{M}}-\rho_{_{R}}\right),
\label{rhoM}\\
\dot\rho_{_{R}} +3(1+\omega_{_R}) H\rho_{_{R}}&=&+\Gamma_M \left(\rho^H_{_M} - \rho_{_{M}}-\rho_{_{R}}\right),
\label{rhoR}
\end{eqnarray}
where $\Gamma_M \left(\rho^H_{_M} - \rho_{_{M}}-\rho_{_{R}}\right)$ represents the interaction between dark energy and matter/radiation via the massive pair plasma state $\rho^H_{_M}$. Equations (\ref{rhoM}) and (\ref{rhoR}) are the cosmic rate equations (\ref{rateeqd}) and (\ref{rateeqr}). 
Here we neglect the decay term $\pm \Gamma_M^{^{\rm de}}\rho^{\rm de}_{_{M}}$ in Eqs.~(\ref{rateeqd},\ref{rateeqr}), assuming unstable massive pairs $\rho^{\rm de}_{_{M}}$ have decayed in reheating. 
The dark-energy equation (\ref{rhoL}) is derived from the generalized Friedman equation (\ref{friedman}) by using cosmic rate equations (\ref{rhoM}) and (\ref{rhoR}). It shows the dark energy interacting with matter and radiation via the back-and-forth 
balance term 
$\Gamma_M(\rho_{_M}^H - \rho_{_{M}}- \rho_{_{R}})$, massive pair plasma state $\rho_{_M}^H$ (\ref{apdenm}) and interacting rate $\Gamma_M$ (\ref{prate}).
% $\Gamma_M^{^{\rm de}}\ll  \Gamma_M$, i.e., $g_{_Y}\ll \chi \epsilon /(4\pi)$. 
%The parameters are $\chi m$ and $g_{_Y}$. The Yukawa coupling $g_{_Y}< \chi \epsilon/(4\pi)$ and $\Gamma_M^{^{\rm de}}< \Gamma_M$.  

During inflation and reheating epochs, the balance term $\Gamma_M \left(\rho^H_{_M} - \rho_{_{M}}-\rho_{_{R}}\right)$ is positive in dark energy equation (\ref{rhoL}) and cosmic rate equations (\ref{rhoM},\ref{rhoR}). 
As results, $\dot\rho_{_\Lambda}<0$ and $\dot\rho_{_{M,R}}>0$, namely dark energy converts to matter and radiation energies \cite{Xue2023,Xue2023a}.  
During standard
cosmology epoch after reheating, two cases are possible: 
\begin{enumerate}[(a)] 
\item negative detailed balance term $\rho^H_{_M}< \rho_{_{M}}+\rho_{_{R}}$, matter and radiation converts to dark energy $\dot\rho_{_\Lambda}> 0$,
\item positive detailed balance term $\rho^H_{_M}>\rho_{_{M}}+\rho_{_{R}}$, dark energy converts to matter and radiation $\dot\rho_{_\Lambda}< 0$. 
\end{enumerate} 
These two cases are separated by $\rho_{_M}+\rho_{_{R}}= \rho^H_{_M}$.

\subsection{A close set of ordinary differential equations for cosmic abundance}\label{mppa}

We define cosmic abundance 
\begin{eqnarray}
\Omega_{_{\Lambda,M,R}}\equiv \frac{\rho_{_{\Lambda,M,R}}}{\rho_{\rm tot}}, \quad \rho_{\rm tot}\equiv \frac{3}{8\pi G}H^2,
\label{abun}
\end{eqnarray}
and $\Omega_{_{\Lambda}}+\Omega_{_{M}}+\Omega_{_{R}}=1$ (\ref{friedman0}). The ``time'' variable $x$ relates to the scale factor $a=a(t)$,
\begin{eqnarray}
x= \ln (a/a_0)+\ln(a_0/a_{_R})= -\ln(1+z) +\ln(a_0/a_{_R}).\label{timex}
\end{eqnarray}
The derivative $dx=Hdt$ and $dx=-dz/(1+z)$, where $z$ is the redshift. Equation (\ref{dde0}) becomes
\begin{eqnarray}
\epsilon= (1+z)\frac{dH}{Hdz}=\frac{3}{2}~\left[(1+\omega_{_M})\Omega_{_M}+ (1+\omega_{_R})\Omega_{_R}+ (1+\omega_{_\Lambda})\Omega_{_\Lambda}\right].
\label{dde0o}
\end{eqnarray}
Whereas, Equations (\ref{rhoL}-\ref{rhoR}) becomes 
\begin{eqnarray}
-(1+z)\frac{d\Omega_{_\Lambda}}{dz}+3(1+\omega_{_\Lambda}) \Omega_{_\Lambda}&=& - 2 \frac{\Gamma_M}{H} \left(\Omega^H_{_M} - \Omega_{_{M}}-\Omega_{_{R}}\right),
\label{rhoLo}\\
-(1+z)\frac{d\Omega_{_M}}{dz} + 3(1+\omega_{_M}) \Omega_{_{M}}&=& +\frac{\Gamma_M}{H} \left(\Omega^H_{_M} - \Omega_{_{M}}-\Omega_{_{R}}\right),
\label{rhoMo}\\
-(1+z)\frac{d\Omega_{_R}}{dz} +3(1+\omega_{_R})\Omega_{_{R}}&=&+\frac{\Gamma_M}{H} \left(\Omega^H_{_M} - \Omega_{_{M}}-\Omega_{_{R}}\right),
\label{rhoRo}
\end{eqnarray}
where $\Omega^H_{_M}=(2/3)\chi(\bar m/M_{\rm pl})^2$ and the $\bar m$ represents the effective mass parameter value $m_{\rm eff}=\bar m$ after reheating. 
The dark energy and matter interacting rate $\Gamma_M/H$ is characterized by the ratio
\begin{eqnarray}
\frac{\Gamma_M}{H}=\frac{\chi \epsilon }{(4\pi)}\frac{(\bar m/H_0)}{(H/H_0)}, 
%\quad \frac{\Gamma_M^{^{\rm de}}}{H}=g_{_Y}\frac{ (m/H_0)}{(H/H_0)}
\label{gammad}
\end{eqnarray}
and $H_0$ is the Hubble constant at the present time $a_0=0$ and $z_0=0$. We define the dark energy and matter exchanging amount $\delta Q$
\begin{eqnarray}
\delta Q\equiv \frac{\Gamma_M}{H} \left(\Omega^H_{_M} - \Omega_{_{M}}-\Omega_{_{R}}\right). 
\label{deltaQ}
\end{eqnarray}
Both rate $\Gamma_M/H$ and amount $\delta Q$ are functions of redshift $z$. These equations are nonlinear, and $\Omega_{_{\Lambda,M,R}}$ and $H$ couple together. 

\subsection{Initial condition at present or reheating time}\label{initial}

The ordinary differential equations (\ref{rhoLo},\ref{rhoMo},\ref{rhoRo}) and condition $\Omega_{_{\Lambda}}+\Omega_{_{M}}+\Omega_{_{R}}=1$ form a closed set. Its solutions $\Omega_{_{\Lambda,M,R}}(z)$ and $H(z)$  
are unique, provided the initial values $\Omega_{_{\Lambda,M,R}}(z_i)$ and $H(z_i)$ are given by measurements at a 
redshift $z_i$. The uniqueness states that the solutions $\Omega_{_{\Lambda,M,R}}(z)$ and $H(z)$ do not depend on the redshift $z_i$ where we implement the initial values. We consider  
initial values at two particular redshifts.
\begin{enumerate}[(i)] 
\item The initial values given by observations today ($a_0=1$ and $z_0(a_0)=0$)
\begin{eqnarray}
\Omega^0_{_{\Lambda}}\approx 0.7,\quad \Omega^0_{_{M}}\approx 0.3,\quad \Omega^0_{_{R}}\approx 3\times 10^{-5},
\label{today}
\end{eqnarray}
and $H_0$. The first-order ordinary differential 
equations (\ref{dde0o}-\ref{rhoRo}) 
determine the solutions $\Omega_{_{\Lambda}}$, $\Omega_{_{M}}$, $\Omega_{_{R}}$ and $H$ as functions of the redshift $z$, varying from the today $z_0=0$ to the past $z_{_R}$ of the reheating end, and to the future $z\rightarrow -1$.
\item The initial values $\Omega_{_{R,M,\Lambda}}(z_{_R})$ and $H(z_{_R})$ are given at the reheating end $a_{_R}$ and 
\begin{eqnarray}
(1+z_{_R})=a_0/a_{_R}\approx (g_*/2)^{1/3}(T_{\rm RH}/T_{\rm CMB}).
\label{zrend}
\end{eqnarray}
The first-order ordinary differential 
equations (\ref{dde0o}-\ref{rhoRo}) 
also determine the solutions $\Omega_{_{\Lambda}}$, $\Omega_{_{M}}$, $\Omega_{_{R}}$ and $H$ as functions of the redshift $z$, varying from the reheating end $z_{_R}$ to the today $z_0=0$, and to the future $z\rightarrow -1$. 
\end{enumerate}
The solution we obtain by using the initial values at $z_0$ must be identical to the one we obtain by using the initial values at $z_{_R}$.
The question is that we do not know values $z_{_R}$, $\Omega_{_{R,M,\Lambda}}(z_{_R})$ and $H(z_{_R})$ at the reheating end $z_{_R}$ (\ref{zrend}), which 
depends on the degeneracy $g^*$ of relativistic particles, the reheating temperature 
$T_{\rm RH}$ and CMB temperature $T_{\rm CMB}$ \cite{Mielczarek2011,Xue2023a}. However, we do know the $z_{_R}\gg 1$ for $T_{\rm RH}\gg T_{\rm CMB}$, and 
the Hubble scale $H(z_{_R})=H_{\rm RH}\sim T_{\rm RH}\gg H_0$. 

\section{Cosmological coincidence problem}\label{coinpro}
In the $\Lambda$CDM model, 
dark energy density 
$\rho_{_\Lambda} = \Lambda/(8 \pi G)$ and $\omega_{_\Lambda}= -1$ are constant in time, decoupling from matter and radiation. This is the case $\Gamma_M\equiv 0$ in Eqs.~(\ref{rhoLo},\ref{rhoMo},\ref{rhoRo}). The ratio of dark energy and matter-radiation components is
\begin{eqnarray}
\frac{\rho_{_{\Lambda}}}{\rho_{_{R}}+\rho_{_{M}}}=\frac{\Omega_{_{\Lambda}}}{\Omega_{_{R}}+\Omega_{_{M}}} \stackrel{\text{$\Lambda$CDM}}{\Longrightarrow} \frac{\Omega^0_{_{\Lambda}}}{\Omega^0_{_{R}}+\Omega^0_{_{M}}}\left(\frac{a}{a_0}\right)^{3\sim 4},
\label{cosmic0}
\end{eqnarray}
which becomes
\begin{eqnarray}
\frac{\Omega_{_{\Lambda}}}{1-\Omega_{_{\Lambda}}}\approx \Omega_{_{\Lambda}}  \stackrel{\text{$\Lambda$CDM}}{\Longrightarrow} 2.33 \left(\frac{a}{a_0}\right)^{3\sim 4}.
\label{cosmic1}
\end{eqnarray}
It shows that at the reheating end $a=a_{_R}\ll a_0$ and $z_{_R}+1=(a_0/a_{_R})\gg 1$, the tiny value $\Lambda\gtrsim 0$ or $\rho_{_{\Lambda}}\gtrsim 0$ must be extremely unnaturally fine-turned many orders of magnitudes to achieve the present ratio 2.33 of dark energy and matter-radiation densities. It is the cosmological coincidence problem of the $\Lambda$CDM, which is caused by (i) $\Lambda$ being a free parameter in usual Friedman equations and (ii) $\rho_{_\Lambda}$ being a constant whereas $\rho_{_{R,M}}\propto a^{-(3\sim 4)}$ decreasing in a very long period of evolution.

In the $\tilde\Lambda$CDM model, basic 
Eqs.~(\ref{rhoL},\ref{rhoMo},\ref{rhoRo}) and initial values (\ref{today}) have uniquely determined the solutions of coupled dark energy and matter-radiation densities and their ratio (\ref{cosmic0}). Likewise, matter/radiation density $\rho_{_{M,R}}$, the dark energy $\rho_{_\Lambda}$ or $\tilde\Lambda$ is a dynamical solution, rather than a free parameter for fine-tuning. However, the issues we need to investigate are
\begin{enumerate}[(i)] 
\item whether or not $\tilde\Lambda$CDM nonlinear Eqs.~(\ref{rhoL},\ref{rhoMo},\ref{rhoRo}) give physical solutions with sensible value $\bar m /H_0$, which evolve in time from the present values (\ref{today}) to the reheating values 
\begin{eqnarray}
\Omega_{_{R}}\gg \Omega_{_{M}}\gg \Omega_{_{\Lambda}} \quad {\rm i.e.,}\quad  \rho_{_{R}}\gg \rho_{_{M}}\gg \rho_{_{\Lambda}},
\label{reheatingend}
\end{eqnarray}
or the equivalent inverse evolution from the reheating end (\ref{reheatingend}) to today (\ref{today}). Namely, it is the unique solutions to Eqs.~(\ref{rhoLo}-\ref{rhoRo}) with boundary conditions (\ref{today}) and (\ref{reheatingend}).  
\item we should obtain such unique solutions without fine-tuning the effective mass parameter $\bar m/H_0$;  
%\red{Otherwise, it would imply that the solutions do not exist if boundary conditions  (\ref{today}) and (\ref{reheatingend}) slightly change};
\item if such solutions exist, how the solutions evolve in an extremely long period after the reheating and achieve the present values (\ref{today}).
\end{enumerate}
Comparing and contrasting with $\Lambda$CDM, we will show how $\tilde\Lambda$CDM solutions possibly evade the cosmological coincidence problem. 

\begin{figure}   
%\begin{center}
%\vspace{-3em}
\includegraphics[height=4.4cm,width=7.0cm]{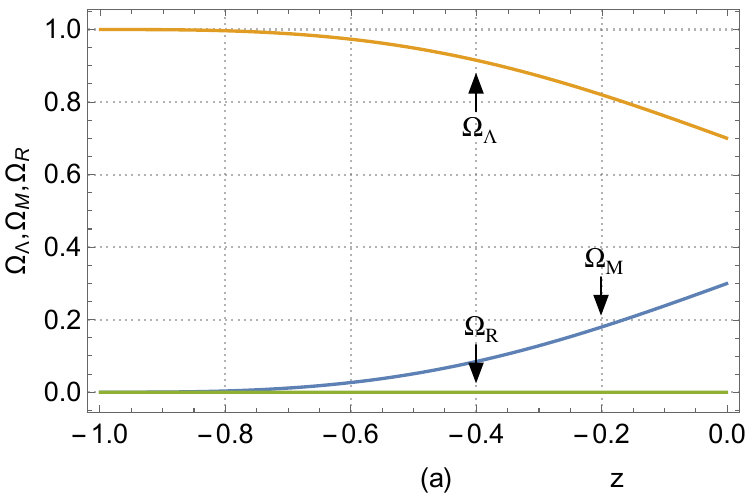}
\includegraphics[height=4.4cm,width=7.0cm]{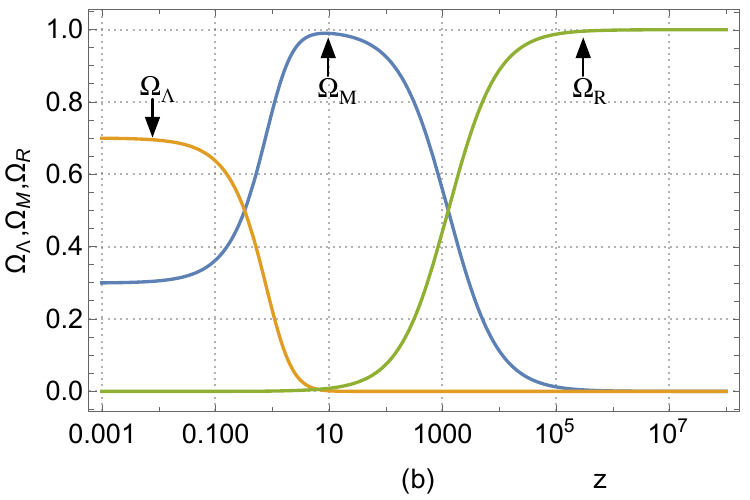}
\includegraphics[height=4.4cm,width=7.0cm]{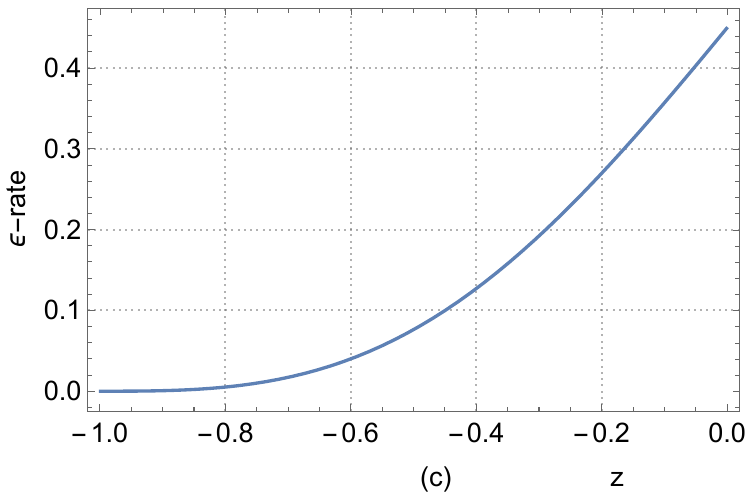}
\includegraphics[height=4.4cm,width=7.0cm]{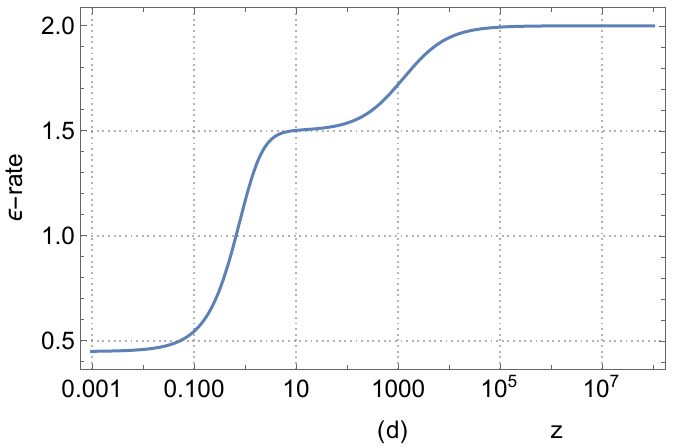}
\includegraphics[height=4.4cm,width=7.0cm]{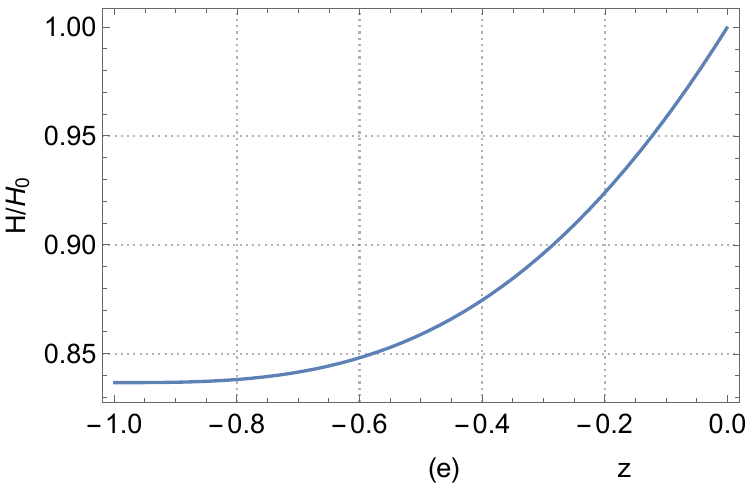}
\includegraphics[height=4.4cm,width=7.0cm]{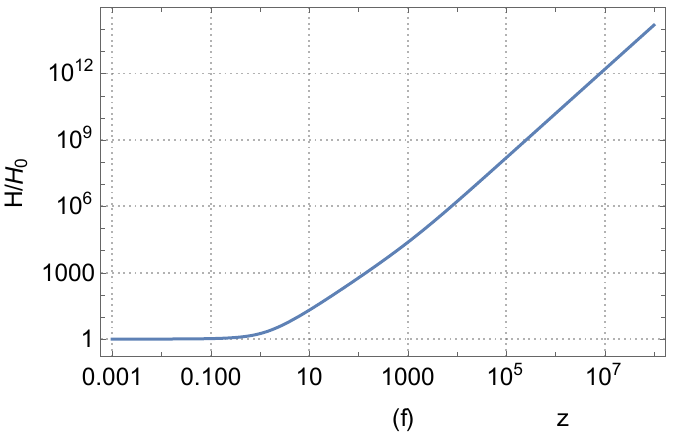}
\includegraphics[height=4.4cm,width=7.0cm]{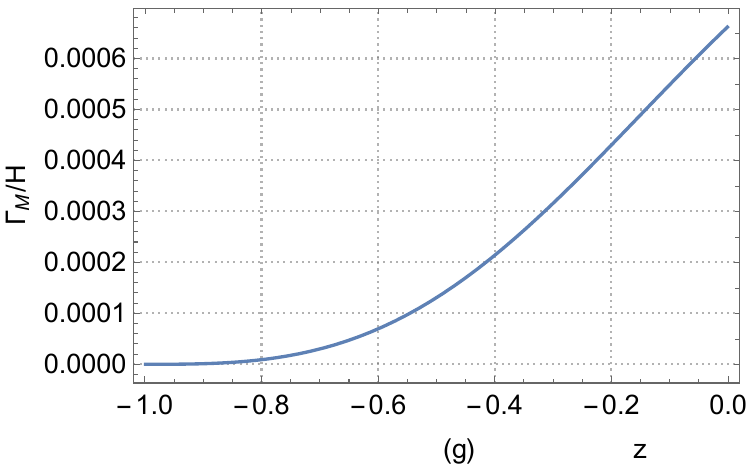}
\includegraphics[height=4.4cm,width=7.0cm]{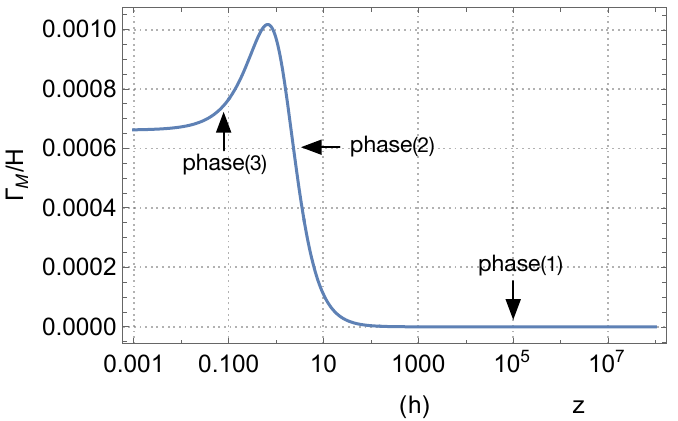}
\includegraphics[height=4.4cm,width=7.0cm]{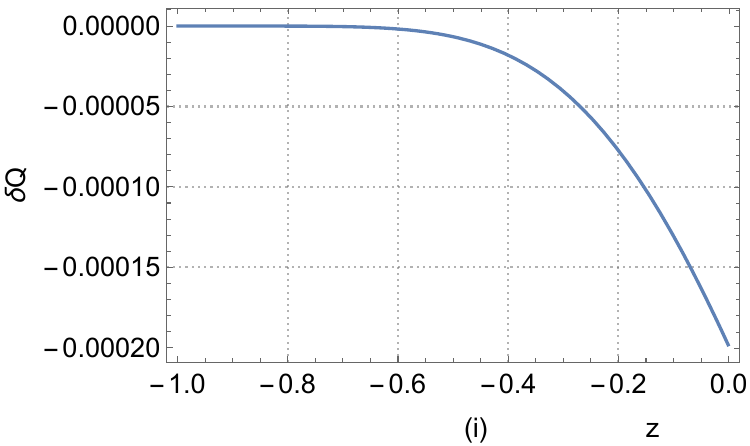}\hskip1.5cm
\includegraphics[height=4.4cm,width=7.0cm]{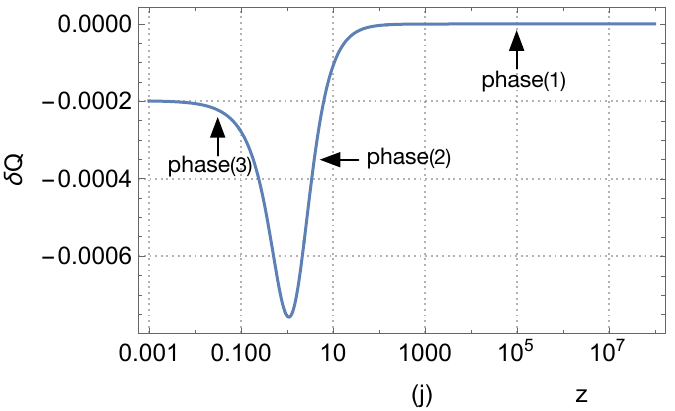}
%\vspace{-1em}
\caption{Numerically solving Eqs.~(\ref{dde0o}-\ref{rhoRo}) and (\ref{today}), $\bar m/H_0=10$, 
%and $\chi=1.85\times 10^{-3}$. 
we present the $\tilde\Lambda$QCD solutions as functions of redshift $z$ from today to the future (the left column) and to the past (the right column). See text for detailed discussions.}
\label{tlcdmplot}
%\end{center}
%\vspace{-2em}
\end{figure}

\subsection{Unique solution to $\tilde\Lambda$CDM equations and boundary conditions}\label{solutiontocoin}

The Hubble scale $H$ 
and scale factor $a$ variations are huge after reheating. The Universe evolves through radiation-, matter- and dark-energy-dominated epochs. The effective values of the mass parameter $\bar m$ can be different in different epochs. %It can also be different for matter and radiation component radiation in eqs. (\ref{rhoMo},\ref{rhoRo}). 
In this article, we study a possible solution to the cosmological coincidence problem, rather than a quantitative analysis of a specific epoch of the Universe.
Therefore, we treat the mass parameter $\bar m/H_0>1$ as an effective mean value for qualitatively studying $\Omega_{_{\Lambda}}$, $\Omega_{_{M}}$, $\Omega_{_{R}}$ and $H$ evolution. The value $\bar m/H_0$ should be fixed by observation data.

We present the numerical solutions in Figure \ref{tlcdmplot}, the left column for the future $0>z>-1$ and the right column for the past $z_{_R}>z>0$. This solution is unique and independent of where the initial condition $z=0$ or $z=z_{_R}$ is implemented. We have numerically checked the uniqueness of the solution. Using the initial values (\ref{today}), we obtain the solution and the final values at $z_{_R}$. As a check, using the final values at $z_{_R}$ as the initial values, we resolve the different equations (\ref{rhoLo},\ref{rhoMo},\ref{rhoRo}) to achieve the values (\ref{today}) at $z_0=0$. 
%We have also numerically checked other initial values differing from (\ref{today}) for the solution being stable .

We first
figure out the novel features of the $\tilde\Lambda$CDM solution, compared and contrasted with the $\Lambda$CDM. The discussions on the numerical solution are in order:
\begin{enumerate}[(i)]
\item Figures \ref{tlcdmplot} (a) and (b) show $\Omega_{_{R, M, \Lambda}}$ evolution in time (or inverse time): (a) from the today $z=0$ (\ref{today}) to the future ($z=-1$) when $\Omega_R\rightarrow 0, \Omega_M\rightarrow 0$ and $ \Omega_\Lambda\rightarrow 1$; 
(b) from $z\approx z_{_R}$ radiation domination $\Omega_R\approx 1, \Omega_M\ll 1$ and $\Omega_\Lambda\approx 0$ to the today $z=0$ (\ref{today}). 
From $z=z_{_R}$ to $z=0$, the matter increases, radiation decreases and $\Omega_R=\Omega_M$ occurs around $z\sim (10^3-10^4)$ for the effective mass parameter $\bar m/H_0\sim (5-10)$. The equality $\Omega_\Lambda=\Omega_M$ occurs around $z\approx 0.2\sim 0.4$, which is not sensitive to the parameter 
$\bar m/H_0$ value. 
Figures \ref{tlcdmplot} (c) and (d) show the Universe evolution $\epsilon$ rate (\ref{dde0}) varies from $\epsilon\approx 2$ 
(radiation) to $\epsilon\approx 3/2$ (matter), $\epsilon \approx 0.45$ (today), and then to $\epsilon\approx 0$ (dark energy) domination. The quantitative results mildly depend on the parameter value $\bar m/H_0$.
All these behaviors 
qualitatively but not quantitatively follow the $\Lambda$CDM model, as shown and explained later. 

\item Figures \ref{tlcdmplot} (e) and (f) show the solution of how the Hubble function in the unit of $H_0$ evolves from $z=z_{_R}$ to $z=0$ then to $z=-1$. The approximate constancy $H\approx H_0$ since $z\approx 0.1$ shows the Universe acceleration and dark energy $\Omega_\Lambda$ domination over matter $\Omega_M$ and radiation $\Omega_R$. Towards the future $z<0$, $H^2\approx (8\pi/3) G\rho_{_\Lambda}$ slowly varies, asymptotically approaching to constant 
$\Omega_\Lambda\lesssim 1$. The evolution is analogous to inflation. 

\item Figures \ref{tlcdmplot} (g) and (h) show the dark energy and matter interacting rate $\Gamma_M/H$ is small and exchanging amount $\delta Q$ (\ref{deltaQ}) is negative. Therefore, the matter and radiation energy slowly convert to the dark energy from the reheating end $z=z_{_R}$ to today $z=0$, then to the future $0 > z> -1$. The dark energy $\Omega_{_{\Lambda}}(z)$ increases 
from $\Omega_{_{\Lambda}}(z_{_R})\approx 0$ 
to $\Omega_{_{\Lambda}}(0)\approx 0.7$ and then to $\Omega_{_{\Lambda}}(-1)\approx 1$. 
For values $\bar m\sim (5\sim 10) H_0$ and $H_0\ll M_{\rm pl}$, we can neglect the $\Omega^H_{_M}=(2/3)\chi(\bar m/M_{\rm pl})^2\ll 1 $, and the $\bar m/H_0$ is a unique parameter for differential equations (\ref{rhoLo}-\ref{rhoRo}). 
\end{enumerate}
We have made numerical verification that the $\tilde\Lambda$CDM 
solution from the maximal $\Omega_{_{M}}$ ($z\approx 5$) to $\Omega_{_{M}}\approx \Omega_{_{\Lambda}}$ ($z\approx 0.5$) 
does not sensitively depend on the effective value of mass parameter $\bar m/H_0$. It implies that the $\tilde\Lambda$CDM solution should not qualitatively change its behaviour, if we adopt the mass parameter $\bar m$ weakly depending 
on $H$.

To end this section, we mention
that in the remote future ($z\rightarrow -1$) of dominant dark energy, radiation and matter $\Omega_R+\Omega_M$ continually decrease until the exchanging amount $\delta Q$ (\ref{deltaQ}) changes sign from negative $\delta Q <0$ to positive $\delta Q >0$. The dark energy density decreases, converting to matter and radiation energy densities. It shows the possibility that the Universe ends the current acceleration and starts recycling again.  
The topic is not the scope of this article. We do not present figures and discussions for this situation in the remote future $z\rightarrow -1$.
%We cannot evidently show in Figures such a situation ($\delta Q >0$) in the  $z\rightarrow -1$, since $\Omega^H_{_M}\ll 1 $ is too small for the parameter $\chi \bar m/H_0$ value chosen.

\subsection{Three evolution phases evading cosmological coincidence problem}\label{three}

The novelty is that the $\tilde\Lambda$CDM results of Fig.~\ref{tlcdmplot} show a natural solution to the cosmic coincidence problem of the $\Lambda$CDM model. %It requires an incredible fine-tuning on the initial value $\Omega_{_{\Lambda}}(z_{_R})\approx 0$ many orders of magnitude, so as to achieve $\Omega_{_{\Lambda}}(0)\sim \Omega_{_{M}}(0)\sim {\mathcal O}(1)$ for an extreme long period from the reheating era $z_{_R}\gg 1$ to the present time $z=0$.
We explain below how the solution works. The results show that the dark energy density almost vanishes ($\rho_{_\Lambda}\gtrsim 0$) at the reheating end $z_{_R}$ and undergoes three evolution phases to achieve its value today.

\begin{figure}   
%\begin{center}
%\vspace{-3em}
\includegraphics[height=4.8cm,width=7.2cm]{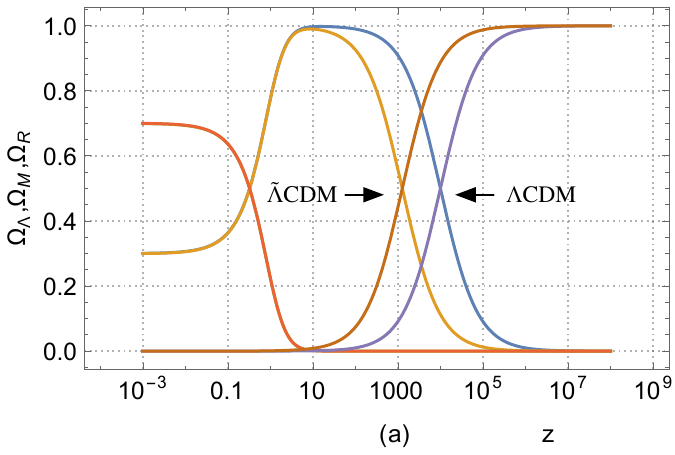}
\includegraphics[height=4.8cm,width=7.2cm]{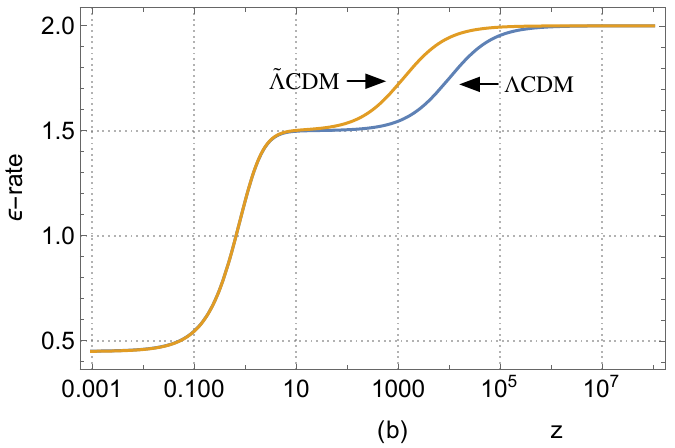}
\includegraphics[height=4.8cm,width=7.2cm]{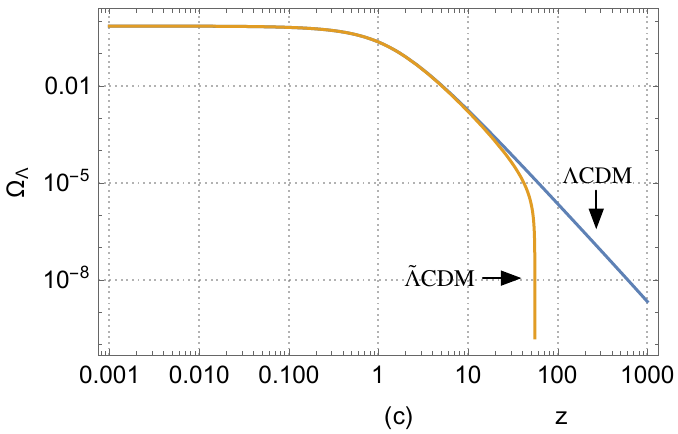}\hskip1.0cm
\includegraphics[height=4.8cm,width=7.2cm]{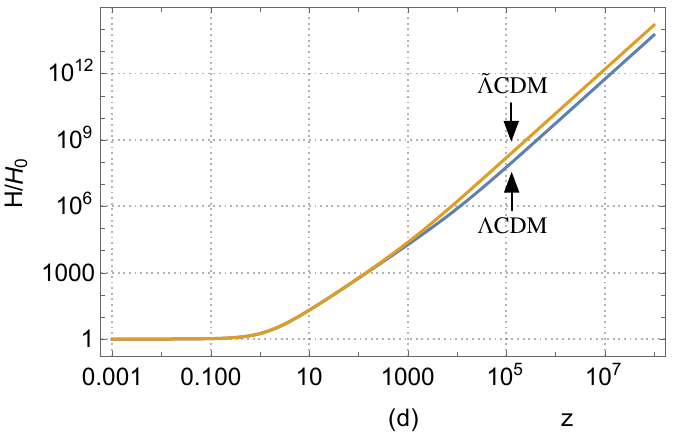}
\includegraphics[height=4.8cm,width=7.2cm]{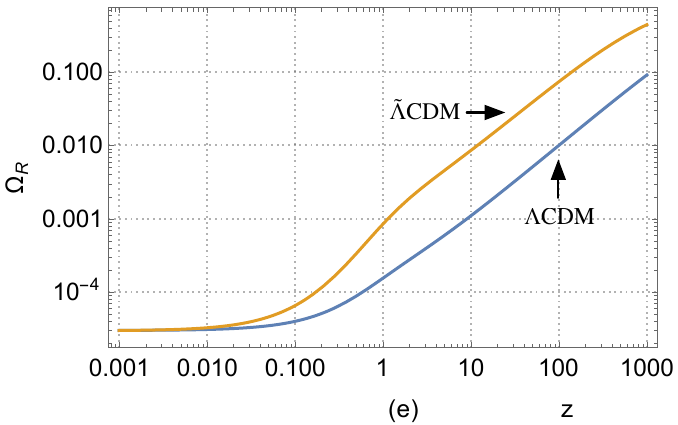}\hskip1.0cm
\includegraphics[height=4.8cm,width=7.2cm]{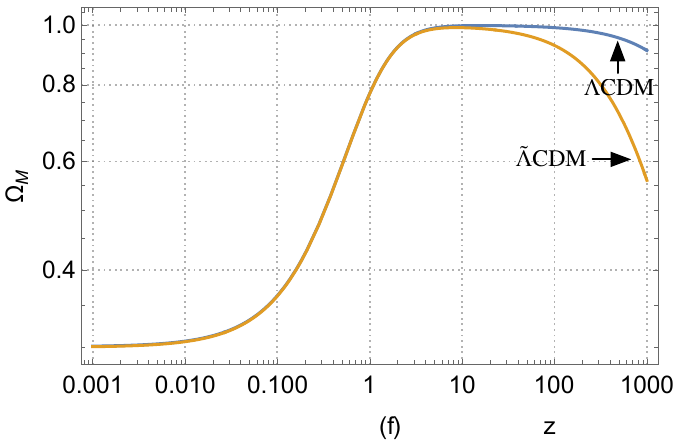}
%\vspace{-1em}
\caption{We present the $\tilde\Lambda$CDM solutions (orange) in comparison and contrast with the canonical $\Lambda$CDM solutions (blue) of Eqs.~(\ref{lcdmd}), \ref{lcdm}) and (\ref{today}). See text for detailed discussions.}
\label{clcdmplot}
%\end{center}
%\vspace{-2em}
\end{figure}

{\it The phase (1):} for a long period ($10\lesssim z\lesssim z_{_R}$), it slowly increases ($\rho_{_\Lambda}\gtrsim 0$) and closely follows up with radiation $\rho_{_R}$ and matter energy $\rho_{_M}$ densities' evolution, see Figures \ref{tlcdmplot} (b) and \ref{clcdmplot} (c) for large $z\gtrsim 10$. The reason is that 
the dark-energy, matter, and radiation interacting rate $\Gamma_M/H\ll 1$ and exchanging amount $\delta Q\ll 1$ are very small, but not zero, see Figure \ref{tlcdmplot} (h) and (j). It is crucial for the dark energy density $\rho_{_\Lambda}$ following up the energy densities $\rho_{_R}$ and $\rho_{_M}$ since they had been varying many orders of magnitudes in the $10\lesssim z\lesssim z_{_R}$ period. This phase is absent in $\Lambda$CDM \footnote{It is a very long period, which is 
why the $\Lambda$CDM has a fine-tuning problem.}. 

{\it The phase (2)}: when the redshift $z\lesssim 10$, the interacting rate $\Gamma_M/H$ and exchanging amount $\delta Q<0$ increase significantly because the Hubble function $H$ becomes smaller and smaller, see Figure \ref{tlcdmplot} (h) and (j). 
The dark energy significantly increases around $z\approx 5$ when the matter $\Omega_M$ domination begins \footnote{This is consistent with the discussions that dark energy evolution follows first radiation then matter in different ways \cite{Xue2019, Xue2020}.}, see Fig.~\ref{tlcdmplot} (b). These features are consistent with the late-time interaction in the dark sector observed by data analysis \cite{Salvatelli2014}.
As a result, in a short period from $z\approx 5$ to $z\approx 0.5$, dark energy $\Omega_{_{\Lambda}}$ increases from $\Omega_{_{\Lambda}}\ll 1$ to the order of unit ${\mathcal O}(1)$. The $\Omega_{_{\Lambda}}$ and $\Omega_{_{M}}$ coincide $\Omega_{_{\Lambda}}\approx\Omega_{_{M}}$ at $z\approx 0.5$. They are in the same order of magnitude up to $z\gtrsim 0$. In this short period $0.5\lesssim z\lesssim 5$, the energy densities $\rho_{_{M}}$ and $\rho_{_{R}}$, and Hubble function $H$ vary only a few orders of magnitudes, see Fig.~\ref{tlcdmplot} (b) and (h), in contrast with their variations in many orders of magnitudes in the long period $10\lesssim z\lesssim z_{_R}$ of the phase (1).
In other words, the low-redshift $\rho_{_{M}}$ and $\rho_{_{\Lambda}}$ evolution are insensible to their initial values at $z_{_R}$. Nature does not need to fine-tune the initial ratios of dark energy, matter and radiation densities at the Big Bang beginning (the reheating end) to achieve $\rho_{_{\Lambda}}/(\rho_{_{M}}+\rho_{_{R}})\sim \mathcal{O}(10^0)$ today. This phase is absent in the $\Lambda$CDM.

{\it The phase (3):} In the period $0\lesssim z \lesssim 0.5$, the dark energy $\Omega_{_{\Lambda}}$, matter and radiation $\Omega_{_{M,R}}$ approach their values today. The 
$\Lambda$CDM has similar behaviours in this phase. As will be shown, the $\Lambda$CDM and $\tilde\Lambda$CDM solutions qualitatively agree with each other.

These three distinct phases (1), (2), and (3) attribute to the properties that the dark energy, matter, and radiation interacting rate $\Gamma_M/H$ (\ref{gammad}) and exchanging amount $\delta Q <0$ (\ref{deltaQ}) are small at high redshift $z$, and large at low redshift $z$. The phase (2) ($z\approx 0.5\sim 10$) separates the phase (3) ($z\approx 0\sim 0.5$) from the phase (1) ($z\approx 10\sim z_{_R}$). Due to such a redshift $z$ dependence of the interaction,
$\tilde\Lambda$CDM evades the problem of fine-tuning $\rho_{_{\Lambda}}$
value at the reheating end to achieve its present value $\rho^0_{_{\Lambda}}$. In other words, it gives a dynamical solution to the cosmic coincidence problem of $\Lambda$CDM without fine-tuning on $\bar m/H_0$. The $\tilde\Lambda$CDM dynamical solution uniquely determines the evolution from today $\Omega_{_{M}}(0)\sim 
\Omega_{_{\Lambda}}(0)\gg \Omega_{_{R}}(0)$ to the reheating end $\Omega_{_{R}}(z_{_R})\gg \Omega_{_{M}}(z_{_R})\gg  \Omega_{_{\Lambda}}(z_{_R})\approx 0$, and {\it vice versa}. Namely, if we would know the initial conditions $\Omega_{_{R, M,\Lambda}}(z_{_R})$ at $z\approx z_{_R}$ (\ref{zrend}), we would have obtained the same dynamical solutions (Fig.~\ref{tlcdmplot}) and the present values (\ref{today}) without fine-tuning. 

To indicate the $\rho_{_\Lambda}$ increase following up $\rho_{_{R, M}}$ evolution in three phases, we here adopt the word ``following-up'' solution, which has a similar sense as the word ``tracker'' solution used in Ref.~\cite{Zlatev1999}. However, the two solutions are different, which we will discuss in Sec.~\ref{distinT}.

We do not discuss the dark energy density perturbations ($\delta\rho_{_\Lambda},\delta p_{_\Lambda}$) caused by its time-varying interaction with matter and radiation. 
However, we speculate that dark energy undergoes transitions and becomes dominant from $z\approx 5$ to $z\approx 0.1$ 
should impact matter density perturbation, leading to the effect on the formation of large-scale structures and clusters. In addition, it should induce the peculiar fluctuations of the gravitational field,  possibly imprinting on observations, for instance, the integrated Sachs-Wolfe effect or galaxy positions. 
The reason is that the dark energy $\Lambda$ results on the gravitation field are very different from the gravitational potential of matter.  

\begin{figure}   
%\begin{center}
%\vspace{-3em}
\includegraphics[height=4.8cm,width=7.2cm]{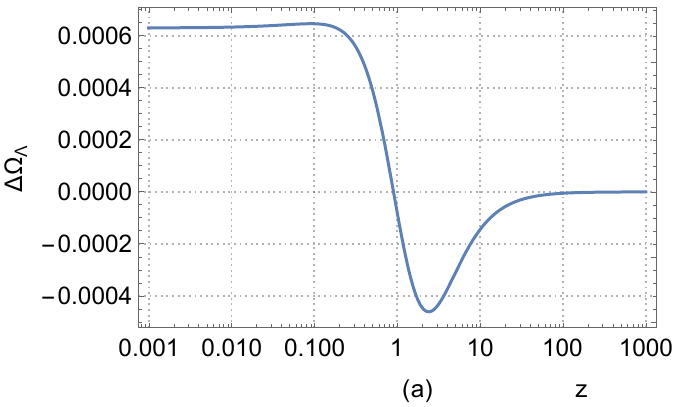}
\includegraphics[height=4.8cm,width=7.2cm]{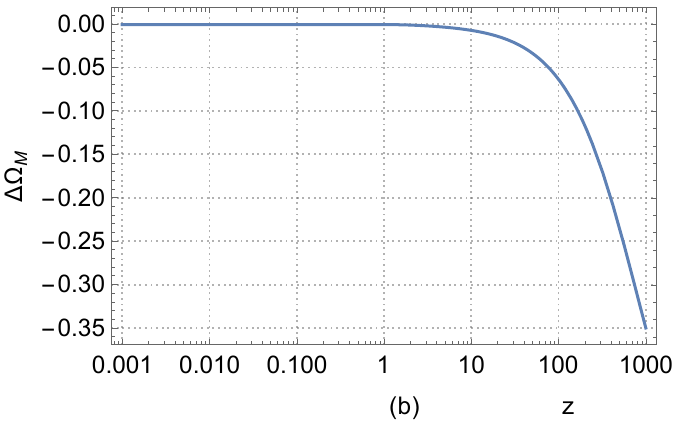}
\includegraphics[height=4.8cm,width=7.2cm]{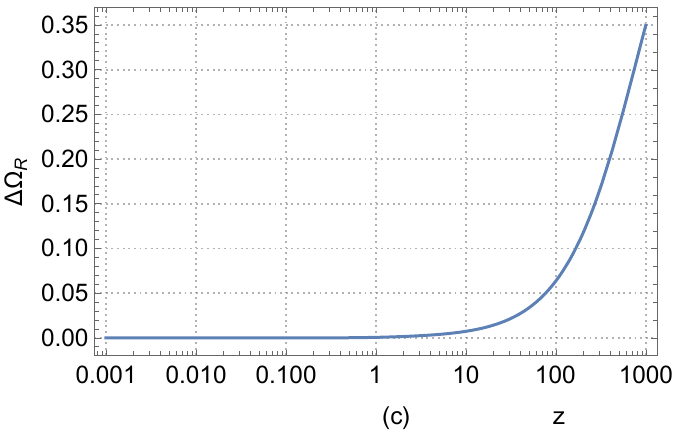}\hskip1.0cm
\includegraphics[height=4.8cm,width=7.2cm]{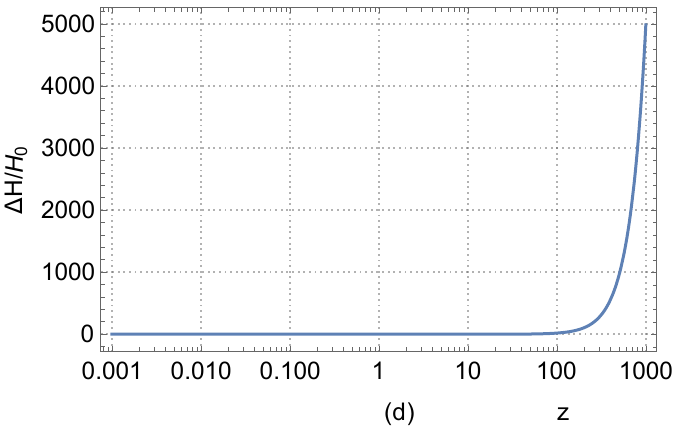}
%\vspace{-1em}
\caption{We present the quantitative differences (\ref{dcdm}) between $\tilde\Lambda$CDM (\ref{today}) and $\Lambda$CDM models (\ref{lcdmd}). See text for detailed discussions.}
\label{plcdmplot}
%\end{center}
%\vspace{-2em}
\end{figure}

\section{Comparison and contrast with $\Lambda$CDM and other models}\label{Repoch}

In this section, we present the $\tilde\Lambda$CDM solutions in comparison and contrast with the $\Lambda$CDM solutions, other dark-energy interacting model solutions, and tracker solutions for cosmic coincidence.

\subsection{Contrast with $\Lambda$CDM solutions}\label{distinL}

In the $\Lambda$CDM model, we define the cosmic abundance of radiation, matter, and dark energy
\begin{eqnarray}
\Omega_{_R}=\Omega^0_{_R}\frac{(1+z)^4}{E(z)^2},\quad
\Omega_{M}=\Omega^0_{_M}\frac{(1+z)^3}{E(z)^2},\quad
\Omega_{_\Lambda}=\Omega^0_{_\Lambda}\frac{(1+z)^0}{E(z)^2},\label{lcdmd}
\end{eqnarray}
and the dimensionless Hubble function $E(z)\equiv (H/H_0)$
\begin{eqnarray}
E(z)^2 =\Omega^0_{_R}(1+z)^4 +\Omega^0_{_M}(1+z)^3+\Omega^0_{_\Lambda}(1+z)^0,
\label{lcdm}
\end{eqnarray}
where $E(0)^2=\Omega^0_{_R}+\Omega^0_{_M}+\Omega^0_{_\Lambda}=1$.
The evolution $\epsilon$-rate is given by Eq.~(\ref{dde0o}).
The values  $\Omega^0_{_R},\Omega^0_{_M}, \Omega^0_{_\Lambda}$ and $H_0$ at $z=0$ are the same as the initial conditions (\ref{today}). Here we use the same notations for quantities of the $\tilde\Lambda$CDM and $\Lambda$CDM models. The former is the dark energy and matter interacting solutions to Eqs.~(\ref{dde0o}-\ref{rhoRo}). The latter is (\ref{lcdmd}) and (\ref{lcdm}) for the constant dark energy density. We have implemented only one observed data point (\ref{today}) for both models.  

\begin{figure}   
%\begin{center}
%\vspace{-3em}
\includegraphics[height=4.8cm,width=7.5cm]{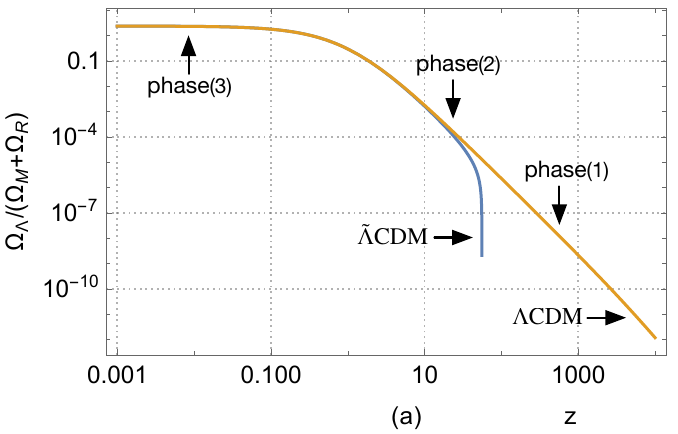}
\includegraphics[height=4.8cm,width=7.5cm]{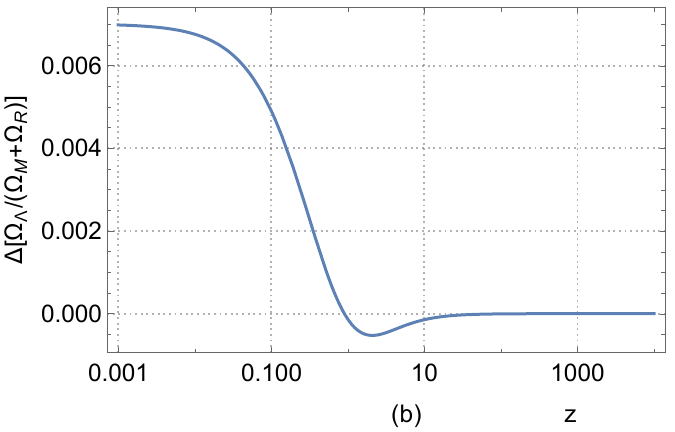}
%\vspace{-1em}
\caption{We present in left the ratio $\rho_{_{\Lambda}}/(\rho_{_{R}}+\rho_{_{M}})$ (\ref{cosmic0}) of dark energy and matter-radiation densities and in right its quantitative differences between $\tilde\Lambda$CDM and $\Lambda$CDM models. In phase (1), the $\tilde\Lambda$CDM ratio is about zero. The dark energy density $\rho_{_{\Lambda}}\ll \rho_{_{R,M}}$ remains 
nearly zero because its interaction with matter and radiation is tiny $\Gamma_M/H\ll 1$.  
In phase (2), the $\tilde\Lambda$CDM ratio rapidly
increases at $z\sim 10$ when $\Omega_M$ approaches to its maximum. The dark energy density closely follows up the matter and radiation densities because the interaction rate 
$\Gamma_M/H$ increases. Namely, the dark energy density runs into the ``tracker solution'' named in the quintessence model. In phase (3), 
the $\tilde\Lambda$CDM ratio is only sensitive to its phase (2) behaviour and insensitive to phase (1). While the $\Lambda$CDM ratio depends on its behaviour in phase (1) up to the reheating end. It explains how $\tilde\Lambda$CDM evades the cosmological coincidence problem of fine-tuning the initial $\rho_{_{\Lambda}}$ value at the reheating end. Although
$\tilde\Lambda$CDM and $\Lambda$CDM ratios approach together, they can be quantitatively different, as shown in the right figure,  Figs.~\ref{clcdmplot} and \ref{plcdmplot}. 
Note that we cannot make plots in the region of small numbers due to the limit of the numerical approach.}
\label{rlcdmplot}
%\end{center}
%\vspace{-2em}
\end{figure}

In Figs.~\ref{clcdmplot} and \ref{plcdmplot}, we compare $\tilde\Lambda$CDM solutions (Fig.~\ref{tlcdmplot}) with the $\Lambda$CDM  (\ref{lcdmd}-\ref{lcdm}) results. The discussions are in order. 
\begin{enumerate}[(i)]
\item Figures \ref{clcdmplot} (a) and (b) show $\Omega_{_{R, M,\Lambda}}$ and expansion rate $\epsilon$ (\ref{dde0o}) evolution for $\Lambda$CDM and $\tilde\Lambda$CDM. Overall, they are consistent and qualitatively agree with each other for $z<10$. 
Two $\Omega_{_{\Lambda}}$ curves overlap in (a), and the point $\Omega_{_{\Lambda}}= \Omega_{_{ M}}$ is about the same, but quantitative numbers are different. The main
differences are (i) the $\epsilon$-rate are in the range $z\sim 10^2\sim 10^5$, (ii) the crossing point $\Omega_{_{R}}= \Omega_{_{ M}}$, and (iii) $\Omega_{_{R, M, \Lambda}}$ in large redshift $z>10$, see Figs.~\ref{clcdmplot} (c), (e), and (f). We show in Fig.~\ref{plcdmplot} the quantitative differences 
\begin{eqnarray}
\Delta\Omega_{_{\Lambda,M,R}}&=&\Omega_{_{\Lambda,M,R}}\big|_{\tilde\Lambda {\rm CDM}}-\Omega_{_{\Lambda,M,R}}\big|_{\Lambda {\rm CDM}},\nonumber\\
\Delta(H/H_0)&=&(H/H_0)\big|_{\tilde\Lambda {\rm CDM}}-(H/H_0)\big|_{\Lambda {\rm CDM}}
\label{dcdm}
\end{eqnarray}
between $\tilde\Lambda$CDM and $\Lambda$CDM.
We have made numerical checks and found these differences mildly depend on the $\tilde\Lambda$CDM parameter value $\bar m/H_0$ in the range ${\mathcal O}(10^{1}) \sim {\mathcal O}(10^{2})$.  

\item Figures \ref{clcdmplot} (d), (e), and (f) show
$H/H_0$, $\Omega_{R}$ and $\Omega_{M}$ evolution for $\tilde\Lambda$CDM and $\Lambda$CDM. Figures \ref{plcdmplot} (b), (c), and (d) show the differences between the two models. 
The discrepancies between $\tilde\Lambda$CDM and $\Lambda$CDM are significant at high red-shift $(z > 10^2)$. These comparisons and contrasts imply that the $\tilde\Lambda$CDM could relieve the $H_0$ and $S_8$ tensions between the values measured today and values calculated in the $\Lambda$CDM model based on measurements at high red-shifts $z$.

\item Figure \ref{clcdmplot} (c) shows
$\Omega_{_{\Lambda}}$ evolution for $\tilde\Lambda$CDM and $\Lambda$CDM.  The crucial difference in $\Omega_{_{\Lambda}}$ appears for $z>10$. 
Due to the nature of constant dark energy density, the $\Lambda$CDM $\Omega_{_{\Lambda}}= \Lambda/(3H^2)$ dependence continues to high redshifts $z\rightarrow z_{_R}\gg 1$ and $H_{\rm RH}\gg H_0$. Therefore, it leads the problem of fine-tuning $\Lambda$ value for achieving $\Omega^0_{_{\Lambda}}= \Lambda/(3H^2_0)\sim {\mathcal O}(1)$ from $\Omega_{_{\Lambda}}= \Lambda/(3H^2_{\rm RH})\ll {\mathcal O}(1)$. However, this is not the case for the $\tilde\Lambda$CDM $\Omega_{_{\Lambda}}= \tilde\Lambda(z)/(3H^2)$, which evolves through phases (1), (2) and (3). The present value $ \Omega^0_{_{\Lambda}}$ mainly depends on variation $\Omega_{_{\Lambda}}(z)$ around small redshift $z\sim 5$ of the phase (2). It is no longer sensitive to fine-tuning value and variation $\Omega_{_{\Lambda}}(z)$ at very large redshifts $z\gg 1$ in phase (1), where $\rho_{_{\Lambda}}(z)$ follows up $\rho_{_{R,M}}(z)$ by weak interaction rate $\Gamma_M/H\ll 1$ (\ref{gammad}) and small exchange $\delta Q\lesssim 0$ (\ref{deltaQ}). Therefore,  $\tilde\Lambda$CDM evades the $\Lambda$CDM fine-tuning problem of cosmological coincidence. We further show in Fig.~\ref{rlcdmplot} three phases (1), (2), and (3) by plotting the dark energy and matter ratio (\ref{cosmic0}) as a function of the redshift $z$, respectively, for $\tilde\Lambda$CDM and $\Lambda$CDM models.

\end{enumerate}

These discussions show that (i) apart from solving the cosmic coincidence problem, the $\tilde\Lambda$CDM's quantities slightly deviate from the $\Lambda$CDM counterparts for $z< 10^3$; (ii) the $\tilde\Lambda$CDM represents a one-parameter $\bar m/H_0$ extension to the $\Lambda$CDM model. 

\subsection{Distinct from other dark-energy interacting models}\label{distin}

We have to point out that dynamical equation (\ref{rhoLo}-\ref{rhoRo}) and $\delta Q$ (\ref{deltaQ}) differ from general modelling dark energy and matter interactions based on total mass-energy conservation and introducing the dark and matter energy's exchange $\delta Q_{\rm int}\propto H\rho$, where $\rho$ relates to energy density, see review \cite{Wang2016, DiValentino2020}. %and \cite{Guo2017, Feng2019, Guo2018}. 
In comparison, we find the crucial differences between the $\tilde\Lambda$CDM and interacting dark energy model: 
\begin{enumerate}[(i)] 
\item the $\tilde\Lambda$CDM interacting term $\delta Q$ (\ref{deltaQ}) changes its sign depending on the dark energy converting to matter or inverse process, while the dark energy interacting model $\delta Q_{\rm int}\propto H\rho$ does not change the sign in evolution;
\item the $\tilde\Lambda$CDM interacting rate (\ref{gammad}) non-linearly relates to $1/H$, see Figs.~\ref{tlcdmplot} (f), (h), and (j) in different phases, while dark energy interacting models' $\delta Q_{\rm int}\propto H$ proportionally. 
\end{enumerate}
The difference (ii) shows that in 
the $\tilde\Lambda$CDM scenario, the dark energy and matter interacting rate is small in the early time of large redshift and becomes large in small redshift for the later time. As shown in Secs.~\ref{solutiontocoin} and \ref{three}, it is an important feature for evading the fine-tuning problem of cosmological coincidence.

\subsection{Comparison with quintessence tracker solutions}\label{distinT}

The comparisons between the $\tilde\Lambda$CDM ``following-up'' solution and quintessence ``tracker'' solution \cite{Zlatev1999} are as follows. Since dark energy density follows below radiation and matter energy densities for most of the history of the universe, both solutions are insensitive to initial conditions (values) at high redshifts and converge to the standard cosmology at low redshifts. The tracker solution differs from the self-adjusting attractor solution approaching a fixed point of autonomous differential equations of dynamical system \cite{Zlatev1999}. Also, the $\tilde\Lambda$CDM solution is not an attractor solution of dynamical equations (\ref{rhoLo}-\ref{rhoRo}).

On the contrary to the $\tilde\Lambda$CDM scenario, there is no particle production for matter or radiation in the quintessence of cosmological scalar field $\phi$, which represents the dark energy component. The decrease in the scalar field $\phi$ potential energy, 
rolling down the potential $V(\phi)$, is converted to its kinetic energy. The tracker solution is a peculiar rolling-down solution, which maintains condition $V''(\phi)\propto H^2$ in the background density of matter and radiation. The ratio $\rho_{_\phi}/(\rho_{_{R}}+\rho_{_{ M}})$ changes steadily. How and where the scalar field $\phi$ runs into the tracker solution depends on the potential $V(\phi)$ and parameters. 
It differs from the $\tilde\Lambda$CDM scenario based on the nontrivial 
redshift dependent interaction (\ref{rhoLo}-\ref{rhoRo}) 
between dark energy and background energy of radiation/matter in phases (1), (2) and (3), as indicated in Figs.~\ref{tlcdmplot} and \ref{rlcdmplot}. 

\subsection{Approximated $\tilde\Lambda$CDM solution for phenomenological studies}\label{appL}

%The dynamic system formed by four differential equations (\ref{dde0o}-\ref{rhoRo}) should have a fixed point at low redshifts ($z\ll 1 $), where the $\Lambda$CDM realizes. From high redshifts ($z\gg1$), the $\tilde\Lambda$CDM quantities approach this fixed point in scaling laws, namely, scaling factors $(1+z)^\delta$ corrected $\Lambda {\rm CDM}$ counterparts. 

Our results base on numerical solutions to $\tilde\Lambda$CDM differential equations (\ref{dde0o}-\ref{rhoRo}). Therefore, it is not convenient in practice for quantitatively comparing the $\tilde\Lambda$CDM numerical solutions with observation data. We look for approximate analytical solutions. 

In redshifts $z < 10^3$, the interacting rate $\Gamma_M/H$ and energy exchange $\delta Q$ are small see Figs.~\ref{tlcdmplot} (h) and (j). We approximate the $\tilde\Lambda$CDM quantities as scaling factors $(1+z)^\delta$ corrected $\Lambda {\rm CDM}$ 
counterparts \footnote{In Ref.~\cite{Xue2015}, we expected these dynamics and approximately derive analytical solutions (\ref{lcdma},\ref{lcdmha}) in the spirit of asymptotic safety of gravitational theories \cite{Weinberg2010}. The view of scaling-law  $(1+z)^\delta$ corrections agrees to the small parameter $\chi \bar m/H_0$ in the dark energy and matter interacting rate (\ref{gammad}).}. 
The scaling indexes $|\delta|\ll 1$, because the $\tilde\Lambda$CDM approaches $\Lambda$CDM, as shown in Fig.~\ref{clcdmplot}.  
Therefore, we approximately decouple Eqs.~(\ref{rhoL}-\ref{rhoR}) into 
\begin{eqnarray}
\dot\rho_{_\Lambda}+0 H\rho_{_\Lambda}&\approx & +\delta_{_\Lambda}H\rho_{_{\Lambda}},
\label{rhoLa}\\
\dot\rho_{_{M}} + 3 H\rho_{_{M}}&\approx & -\delta^G_{_R}H\rho_{_{M}},
\label{rhoMa}\\
\dot\rho_{_{R}} +4 H\rho_{_{R}}&\approx &-\delta^G_{_M}H\rho_{_{R}}.
\label{rhoRa}
\end{eqnarray}
Three new dimensionless parameters $\delta^G_{_R}$,  $\delta^G_{_M}$ and $\delta_{_\Lambda}$ are proportional to $\chi \bar m/H_0$, and much smaller than the unity. Equations (\ref{rhoLa}-\ref{rhoRa}) yield the effectively corrected densities    
\begin{eqnarray}
\rho_{_R}\approx \rho^0_{_R}(1+z)^{4-\delta^{^R}_{G}},\quad
\rho_{_M}\approx \rho^0_{_M}(1+z)^{3-\delta^{^M}_{G}},\quad
\rho_{_\Lambda}\approx \rho^0_{_\Lambda}(1+z)^{\delta_{_\Lambda}},\label{lcdma}
\end{eqnarray}
and the Hubble function
\begin{eqnarray}
E^2(z)=\Omega^0_{_R}(1+z)^{4-\delta^{^R}_{G}}+\Omega^0_{_M}(1+z)^{3-\delta_G^{^M}}+\Omega^0_{_\Lambda}(1+z)^{\delta_{_\Lambda}}.\label{lcdmha}
\end{eqnarray}
The equation (\ref{dde0}) or (\ref{friedman0},\ref{friedman}) gives the constraint of parameters 
$\delta^G_{_R}$, $\delta^G_{_M}$ and $\delta_{_\Lambda}$,
\begin{eqnarray}
\delta_\Lambda &\approx & (\Omega^0_{M}\delta^{^M}_{ G}+\Omega^0_{R}\delta^{^R}_{G})/\Omega^0_{_\Lambda},\
 \label{deltal+}
\end{eqnarray}
and two parameters are independent.
In the view of Eqs.~(\ref{lcdma}), we find the equations of states effectively modify: $\omega^{\rm eff}_{_R}\approx (1/3)(1-\delta^{^R}_{G})$, $\omega^{\rm eff}_{_M}\approx -(1/3)\delta^{^M}_{G}$ and $\omega^{\rm eff}_{_\Lambda}\approx -1+(1/3)\delta_{_\Lambda}$, see also Ref.~\cite{Begue2019}.
Equations  (\ref{lcdma},\ref{lcdmha},\ref{deltal+}) are $\tilde\Lambda$CDM approximate solutions, which facilitate data analysis for comparing $\tilde\Lambda$CDM with observational data. 

References \cite{Gao2021, Gao2022} presents detailed numerical studies and data analysis based on the approximated $\tilde\Lambda$CDM solutions (\ref{lcdma},\ref{lcdmha},\ref{deltal+}) and numerous data sets of observations. It shows that both $\Lambda$CDM $H_0$ and $S_8$ tensions reduces to $2\sigma$ level with constraint parameters $\delta^{^R}_{ G}\approx - 1.5\times 10^{-2}$, $\delta^{^M}_{ G}\approx  - 5.0\times 10^{-4}$ and $\delta_{_\Lambda}\approx  - 2.0\times 10^{-4}$. The negative parameter values support the scenario of energy conversion from radiation and matter to dark energy, as discussed in the previous section. The negative $\delta_{_\Lambda}\lesssim 0$ implies that due to interactions, dark energy slightly behaves as if it was a phantom energy $\omega^{\rm eff}_{_\Lambda}\approx -1+(1/3)\delta_{_\Lambda}\lesssim -1$. It differs from the situation in inflation and reheating when dark energy converts to matter and radiation energies \cite{Xue2023a} see Fig.~\ref{clcdmreh}, dark energy behaves as if it was a quintessence energy $\omega^{\rm eff}_{_\Lambda}> -1$. 
%it is then non-phantom energy $\omega_{_\Lambda}\approx -1+(1/3)\delta_{_\Lambda}\gtrsim -1$ for $\delta_{_\Lambda}\gtrsim 0.$ 

\section{Discussions on Einstein cosmological $\Lambda$ term }

We end this article by recalling how $\tilde\Lambda$CDM inflation and reheating processes arrive at the condition $\rho_{_{R}}\gg \rho_{_{M}}\gg \rho_{_{\Lambda}}$ (\ref{reheatingend}) at the reheating end. We have studied this issue in Refs.~\cite{Xue2023, Xue2023a}. Massive pairs' production slows down inflation driven by dark energy, and their decay to relativistic particles leads to the condition (\ref{reheatingend}). 

In addition, we present some speculations on the gravitational (geometric) and dynamical natures of the cosmological $\tilde\Lambda$ term and dark-energy density $\rho_{_\Lambda}=\tilde\Lambda/(8\pi G)$ of the Einstein theory. Then, we discuss the dynamical solution to the cosmic fine-tuning problem.

\subsection{Geometric nature of $\tilde\Lambda$ dark energy as gravitational ground state}\label{geom}
The $\tilde\Lambda$ term possibly represents \cite{Coleman1988, Barvinsky2007, Xue2010, Xue2009, Xue2012,Xue2015}
%, Wang2020a, Wang2020} 
the non-trivial ground state (Wheeler spacetime foam \cite{Misner1973, Carlip2022}) of the spacetime. 
The perturbative quantum gravitational field fluctuates upon such a ground state, and the classical gravitational field varies in such a ground state. They are effectively described by the gravitation coupling $G$, the $\tilde \Lambda$, and the Ricci scalar $R$ terms in Einstein's theory. Such a ground state is probably a coherent state of the long-ranged holonomy field (see Eq.~(133) of Ref.~\cite{Xue2010}). It is a condensate state due to violent quantum gravity at the Planck scale. The spacetime foam structure of such a ground state is most intriguing. It could be an interacting gas of gravitational instantons (wormholes), whose effective equation of state behaves as $p_{_\Lambda}= -\rho_{_\Lambda}$, see Sec.~X of Ref.~\cite{Xue2009a}. We are proceeding with further studies on these aspects. 

The $\tilde\Lambda$ ($\xi\sim 1/\tilde\Lambda^{1/2}$) is the characteristic scale (correlation length) of such nontrivial geometric ground state \cite{Xue2012, Xue2015}. It represents the intrinsic scale for effective gravitational field theories 
realized in the scaling domains of fixed points of effective gravitational coupling $g\sim GM^2_{\rm pl}$ to matter and radiation. The $\tilde\Lambda$ and $g$ varying from one fixed point to another render its dynamic nature. It is nontrivial to demonstrate these dynamical features. However, as analogies, we mention fundamental field theories of interactions (i) the electroweak scale $v\sim 10^2$GeV for electroweak field theory realized in the scaling domain of infrared (IR) fixed point; (ii) the 
scale $\Lambda_{\rm QCD}\sim 10^2$MeV for perturbative QCD field theory realized in the scaling domain of ultraviolet (UV) fixed point; (iii) the low-energy hadron scale for non-perturbative QCD field theory realized in the scaling domain of IR fixed point. 

\subsection{Asymptotically safe Einstein theory for early and present Universe}

On the one hand, in early Universe of $\tilde\Lambda$ dark energy dominated inflation, $H^2\sim (8\pi G/3)\rho_{_\Lambda}
$ and $\rho_{_\Lambda}=\tilde\Lambda/(8\pi G)$ asymptotically give $\xi\sim 1/\tilde\Lambda^{1/2}\sim H^{-1}$. Namely, the correlation length $\xi$ is the size of the horizon. The $\tilde\Lambda^{1/2}$ slowly varies from the inflation scale $H^*$ to the scale $H_{\rm end}\approx (0.42, 0.35)H_*$ at inflation end $a_{\rm end}$.
The $H_*\sim 10^{-6}M_{\rm pl}$ is obtained from the CMB data, see Eqs.~(6.5) and (6.10) of Ref.~\cite{Xue2023}. The inflation scale $H_*$ is much smaller than the Planck scale. How quantum gravitation field theory with the intrinsic scale $\tilde\Lambda^{1/2} \sim M_{\rm pl}$ runs to the effective Einstein theory at the scale $\tilde\Lambda^{1/2}\sim H_*\ll M_{\rm pl}$. How does the quantum gravity ground state evolve to the $\tilde\Lambda$ ground state of effective Einstein theory? One could study it in the context of the asymptotically safe and effective theories of gravitation \cite{Weinberg2010} and the scaling domain of a UV unstable fixed point \cite{Xue2015}. 

On the other hand, in the recent Universe of 
$\tilde\Lambda$ dark energy dominated acceleration, 
$H^2_0\sim (8\pi G/3)\rho^0_{_\Lambda}
$ asymptotically gives the scale $\xi\sim 1/\tilde \Lambda^{1/2}\sim H^{-1}_0$ and density $\rho^0_{_\Lambda}\approx H_0^2/(8\pi G)$  \cite{Gurzadyan2003}. 
Based on the same spirit of asymptotic safety of effective gravitational theories \cite{Weinberg2010}, 
we study its realization in the scaling domain of a UV stable fixed point, where is the effective Einstein theory of relevant operators $R/G$ and $\tilde\Lambda/G$, and gravitational 
coupling $G$ and cosmological $\tilde \Lambda$ approach their 
values today \cite{Xue2012, Xue2015}.  However, due to the dark energy, radiation and matter interactions, as well as pair production of massive particles and antiparticles on the horizon, it is nontrivial to find the scaling laws for operators $R/G$  and $\tilde\Lambda/G$ by using the asymptotic safety principle. 
The questions are how the $\tilde\Lambda$ dark energy varies from the inflation scale $H_*$ to the recent Hubble scale $H_0\ll H_*$. How the dark energy density changes from $\rho^*_{_\Lambda}\approx H_*^2/(8\pi G)$ to $\rho^0_{_\Lambda}\ll \rho^*_{_\Lambda}$ in many orders of magnitudes. We use the $\tilde\Lambda$CDM solutions in inflation, reheating and standard cosmology to explain the possible solution to such cosmic fine-tuning problem.

\begin{figure}   
\begin{center}
%\vspace{-3em}
\includegraphics[height=4.8cm,width=7.2cm]{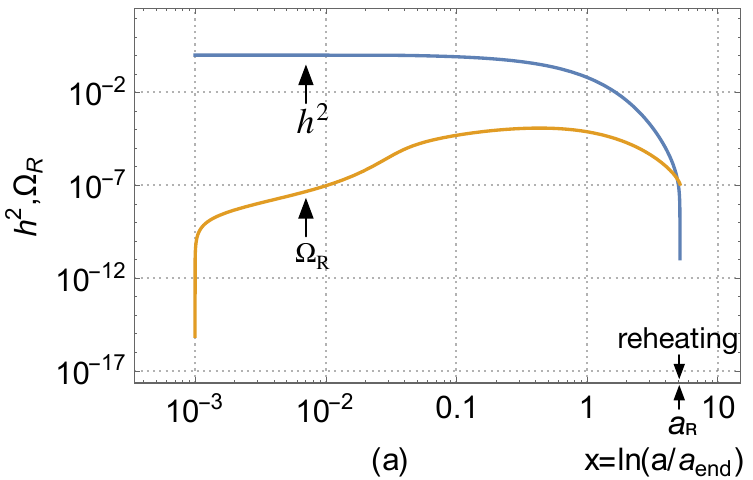}
\includegraphics[height=4.8cm,width=7.2cm]{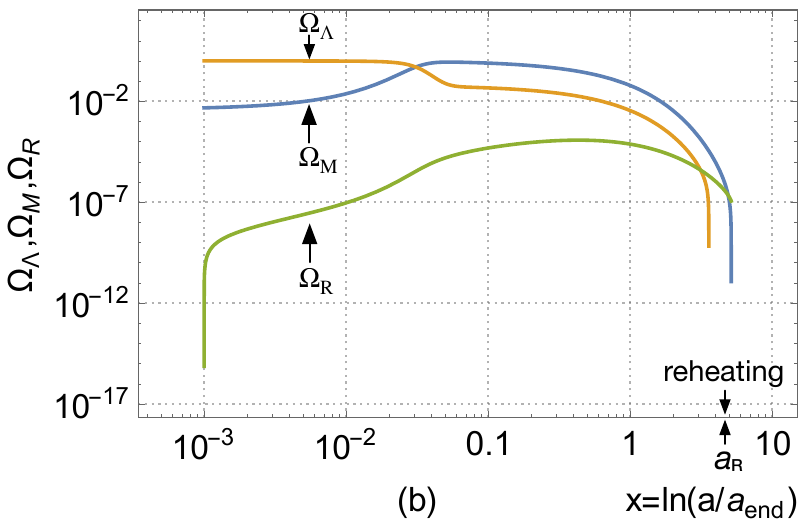}
%\vspace{-1em}
\caption{We reproduce the $\tilde\Lambda$CDM results of the Hubble function $h^2=H^2/H^2_{\rm end}$ (left), $\Omega_M$, $\Omega_R$ and $\Omega_\Lambda$ (right) in reheating, see Figure 6 (a) and (b) in Ref.~\cite{Xue2023a}. They are in the unit of $\rho^{\rm end}_c=3 m^2_{\rm pl}H^2_{\rm end}$ and the Hubble scale $H_{\rm end}\approx (0.42, 0.35)H_*$ at inflation end $a_{\rm end}$, see Eq.~(6.10) of Ref.~\cite{Xue2023}. The parameters $\hat m/m_{\rm pl}\approx 27.7$ and $\chi\approx 1.85\times 10^{-3}$. 
It shows that 
in a few numbers $x=\ln(a/a_{\rm end})$, dark energy $\Omega_\Lambda$ decreases from one to zero because of rapidly converting to matter and radiation. Matter $\Omega_M$ increases and dominates over $\Omega_\Lambda$. 
Then $\Omega_M$ decreases because of decaying to radiation $\Omega_R$. As a result, $\Omega_R$ increases and becomes dominant. At the reheating end $(a_{_R}/a_0)=(1+z_{_R})^{-1}$, the radiation abundance is about one, and the dark energy and matter abundances are about zero.}
\label{clcdmreh}
\end{center}
%\vspace{-2em}
\end{figure}

\subsection{Dynamical nature of $\tilde\Lambda$ dark energy solving fine-tuning problem}

After the inflation ends, the Universe undergoes reheating. Based on dynamical equations (\ref{dde0o}-\ref{rhoRo}), we show \cite{Xue2023a} that due to strong coupling ($\Gamma_M/H\gg 1$) between $\tilde\Lambda$ dark energy and matter energy densities,
dark energy rapidly converts into 
massive matter, and the latter decays to radiation energy. As a result, dark energy density decreases from $\rho^{\rm end}_{_\Lambda}\approx 3m^2_{\rm pl}H^2_{\rm end}$ 
to $\rho^R_{_\Lambda}\approx 0$, where $\rho^R_{_{\Lambda, M, R}}$ stand for the dark energy, matter and radiation densities at the reheating end $a_{_R}/a_0=(1+z_{_R})^{-1}$. 
We illustrate in Fig.~\ref{clcdmreh} the dynamical reheating process from the inflation end $\rho^{\rm end}_{_\Lambda}\gg \rho^{\rm end}_{_M}\gg \rho^{\rm end}_{_R}\approx 0$ to the reheating end $\rho^R_{_R}\gg \rho^R_{_M}\gg \rho^R_{_\Lambda}\approx 0$. The radiation energy density $\rho^R_{_R}$ becomes dominant, initiating the standard cosmology. There is no fine-tuning in this process. 

Then how the standard cosmology dynamically evolves to the coincidence $\rho^0_{_\Lambda}\sim \rho^0_{_M}\gg \rho^0_{_R}\approx 0$ in the recent epoch.
It is the issue addressed in this article. The initial values of the scale factor $a_{_R}$, Hubble constant $H_{\rm RH}$ and energy densities $\rho^R_{_{R,M,\Lambda}}$ cannot be completely determined. Therefore, we cannot uniquely solve ordinary differential equations (\ref{dde0o}-\ref{rhoRo}) from the reheating end $z_{_R}$ to the present epoch $z=0$. However, we use the present values (\ref{today}) to uniquely solve ordinary differential equations (\ref{dde0o}-\ref{rhoRo}) from today $z=0$ back to the reheating end $z_{_R}\gg 0$.  As shown in Fig.~\ref{tlcdmplot} (b) and Fig.~\ref{clcdmplot} (c), the dynamical solutions asymptotically approach the same initial conditions $\rho^R_{_R}\gg \rho^R_{_M}\gg \rho^R_{_\Lambda}\approx 0$ for $z\rightarrow z_{_R}\gg 1$ without any fine-tuning. 

Such qualitative matching implies a consistent dynamical solution for the cosmic fine-tuning problem in the following way. Converting to matter and radiation $\delta Q\gg 1$ (\ref{deltaQ}), the dark energy density
decreases from the inflation scale $\rho^*_{_\Lambda}\approx 3m^2_{\rm pl}H^2_*$ to inflation end $\rho^{\rm end}_{_\Lambda}\approx 3m^2_{\rm pl}H^2_{\rm end}$, then to reheating end $\rho^R_{_\Lambda}\approx 0$. Since the standard cosmology starts, converted from matter and radiation $\delta Q\lesssim 0$ (\ref{deltaQ}), the dark energy density increases from the reheating end value $\rho_{_\Lambda}^R\approx 0$ to the present value $\rho^0_{_\Lambda}\approx H_0^2/(8\pi G)$ \cite{Gurzadyan2003}.  
Such a dynamical evolution is free from fine-tuning. It can be the solution to the cosmic coincidence problem. The basic reasons are that 
in evolution dark energy and matter conversion $\delta Q$ (\ref{deltaQ}) 
changes sign and is proportional to the interacting rate $\Gamma_M/H\propto \chi m \epsilon/H$ and $m>H$. 

Nonetheless, we have not yet found the complete and quantitative solution to the cosmic fine-tuning problem since we separately adopt the effective values of mass parameter $m_*/H_*$ for inflation, $\hat m/H_{\rm end}$ for reheating and $\bar m/H_0$ for standard cosmology. The mass parameter $m$ is proportional to the mass $M$ and number ${\mathcal N}_{\rm pair}$ of massive particle
and antiparticle pairs in the holographic massive plasma state. Therefore, its value should depend on the horizon $H$, i.e., $m=m(H)$. Its time-varying should be slower than the $H$ so that the dark energy and matter interacting rate $\Gamma_M/H\propto m(H)/H\propto m_{\rm eff}/H$ decreases (increase) as $H$ increases (decreases).   
We have not been able to determine $m(H)$ in theory. Instead, we define its effective values $m_{\rm eff}$ of each epoch, for instance, $m^R_{\rm eff}$ for radiation domination and $m^M_{\rm eff}$ for matter domination, whose values should be fixed by observations. 
To understand these fundamental natures of the Einstein cosmological $\Lambda$ term in cosmology, further observations of the cosmos are necessary \cite{Amendola2013, Amendola2018}.

%\vskip0.1cm
%\section{\bf Acknowledgment}
%\red{The author also thanks the anonymous referees for their reports that improve the article.} 
%\newpage

\section{Appendix: quantum pair oscillation details}\label{qppos}

We add this appendix for readers' convenience to gain a preliminary picture of massive pair productions and oscillations without going into details of Ref.~\cite{Xue2023, Xue2023a}. We show the massive particle production and oscillation coupling with the Hubble function fast component $H_{\rm fast}$. For a given $H$ and $h^{\rm fast}=H_{\rm fast}/H$,
we express in unit of the critical density $\rho_{\rm crit}=3m_{\rm pl}^2H^2$ the dimensionless quantum pressure ${\mathcal P}_{_M}^{\rm fast}$ and density 
$\varrho^{\rm fast}_{_M}$.  In dimensionless microscopic time $t$ in the unit of $M^{-1}$, we plot the Bogoliubov coefficient $|\beta|^2$, the quantum pair density  $\varrho^{\rm fast}_{_M}$ and pressure ${\mathcal P}_{_M}^{\rm fast}$, as well as the fast components of the Hubble function $h_{\rm fast}$, 
%\red{remove: scale factor $a_{\rm fast}$} 
and cosmological term
$\varrho^{\rm fast}_{_\Lambda}$. 
The details and notations are given in Secs.~2-3 of Ref.~\cite{Xue2023a}. 

In Figs.~1, 3 and 4 of Ref.~\cite{Xue2023}, we show the features of dynamical oscillation in pre-inflation and inflation 
epochs for the mass parameter $m_*\approx 3.1 m_{\rm pl}$. In Figs.~1 and 11 of Ref.~\cite{Xue2023a}, we show the features of dynamical oscillation reheating epoch for the parameter $\hat m \approx 20 m_{\rm pl}$. While for the standard cosmology after reheating, both the Hubble scale $H$ and pair mass $M$ are very much smaller than the Planck mass, i.e., $H\ll M \ll m_{\rm pl}$ and pair number ${\mathcal N}_{\rm pair}\gg 1$. Therefore, we present these plots for the selected values $H/M$ and ${\mathcal N}_{\rm pair}$, which differ from 
those in inflation and reheating. It 
also gives an insight into the mass parameter $m$ variation from one epoch to another. 

\begin{figure*}[t]
\centering
\begin{center}
%\vspace{-1.5em}
\includegraphics[height=5.5cm,width=7.8cm]{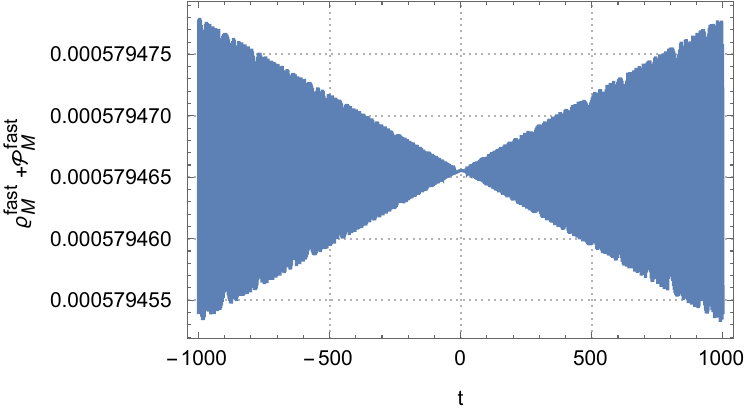}
\caption{We show the quantum pair density and pressure oscillations in microscopic time $t$ in the unit of $M^{-1}$, using $H/M 
\approx 10^{-3}$, $M\simeq 10^{-10}m_{\rm pl}$, 
${\mathcal N}_{\rm pair}\simeq 2\times 10^{22}$ and $\delta= 1$. 
It shows that a large number of massive pairs 
creates significant oscillating quantum pressure ${\mathcal P}^{\rm fast}_{_M}$ and $\varrho^{\rm fast}_{_M}$  
in the unit of $\rho_{\rm crit}=3m_{\rm pl}^2H^2$. In this coherent state, ${\mathcal P}^{\rm fast}_{_M}$ and $\varrho^{\rm fast}_{_M}$ oscillation 
frequency is about $M$. Their oscillating amplitudes $\delta\varrho^{\rm fast}_{_M}/\varrho^{\rm fast}_{_M}$ and $\delta {\mathcal P}^{\rm fast}_{_M}/{\mathcal P}^{\rm fast}_{_M}$ are about ${\mathcal O}(10^{-3})$. For a long time, %$t> 10^4$, 
the coherent oscillations approach stable configurations in time. % and amplitude-damping effects appear. 
Note that the pair number ${\mathcal N}_{\rm pair}$, mass scales $M$ and $H$ values differ from those used for inflation see Fig.~1 in Ref. \cite{Xue2023} and reheating see Fig.~1 in Ref.~ \cite{Xue2023a}.}
\label{osci+}
\end{center}
\vspace{-2em}
\end{figure*}

\begin{figure*}[t]
%\centering
%\begin{center}
\vspace{+3em}
\includegraphics[height=5.5cm,width=7.8cm]{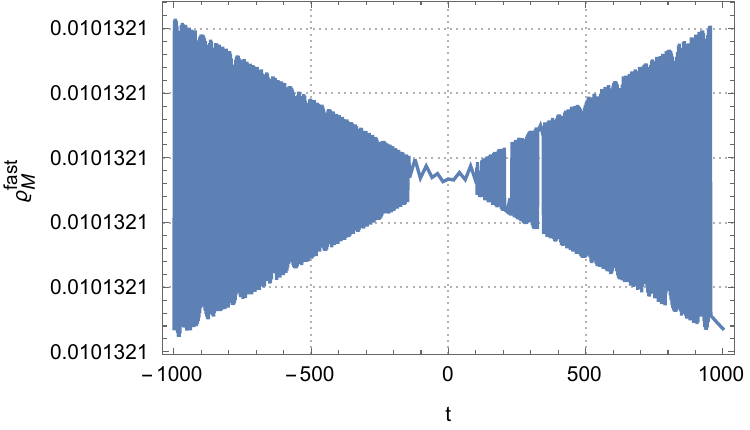}\hspace{0.333cm}
\includegraphics[height=5.5cm,width=7.8cm]{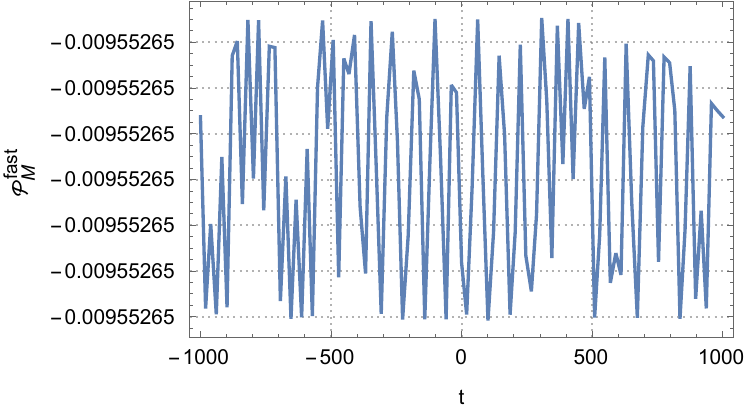}\hspace{0.333cm}
\includegraphics[height=5.5cm,width=7.8cm]{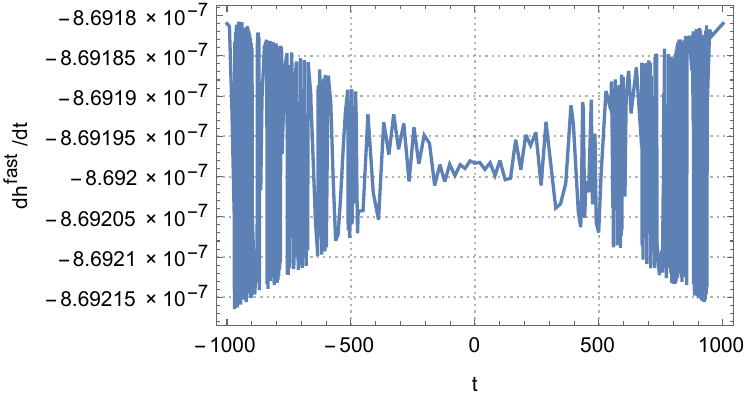}\hspace{0.333cm}
\includegraphics[height=5.5cm,width=7.8cm]{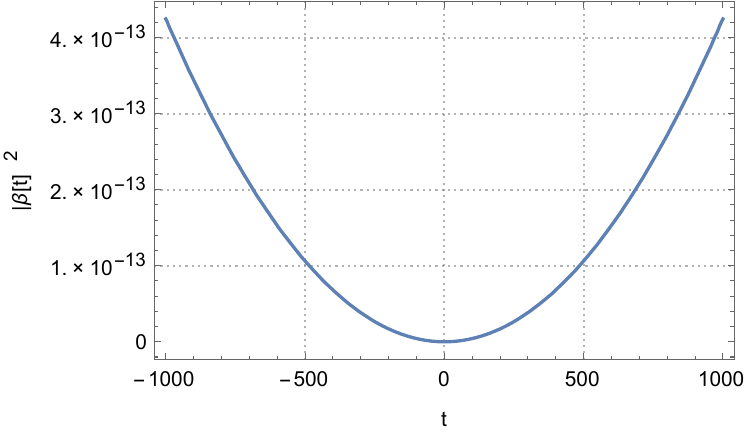}
\includegraphics[height=5.5cm,width=7.8cm]{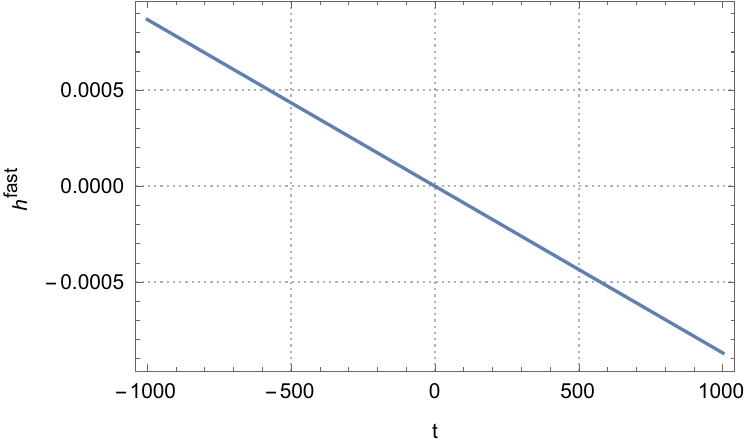}\hspace{0.333cm}
\includegraphics[height=5.5cm,width=7.8cm]{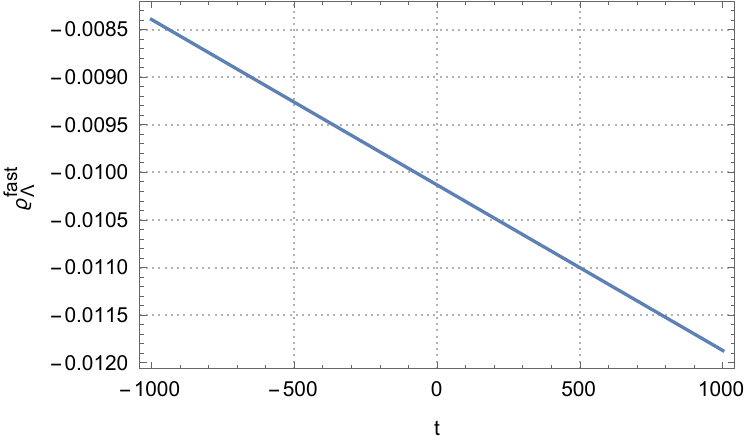}
\caption{Corresponding to Fig.~\ref{osci+}, we show the details of quantum pair oscillation in microscopic time $t$ 
in the unit of $M^{-1}$. The oscillatory $|\beta(t)|^2$, 
$h_{\rm fast}=H_{\rm fast}/H$ and $\varrho^{\rm fast}_{_\Lambda}$ structures are too small to see. 
The parameters' values are the same as those in Fig.~\ref{osci+}. The non-smooth curve $|\beta(t)|^2$ shows its oscillating behaviour. The $h_{\rm fast}$ and $\varrho^{\rm fast}_{_\Lambda}$ oscillatory structures are too small to see due to the precision limit for numerical calculations with the parameters' values used. However, one can infer their oscillating behaviours by the oscillating $dh^{\rm fast}/dt$. We suggest readers see Fig.~4 of Ref.~\cite{Xue2023} for pre-inflation and inflation, where the corresponding solutions for other parameters' values and plotting scales show evident oscillatory structures.}
\label{detailosci1+}
%\end{center}
%\vspace{-2em}
\end{figure*}

%\bibliography{OscillationRef}
%\bibliography{../OscillationRef}
%\bibliography{../../MyBibFiles/MyBibFile.bib}

\providecommand{\href}[2]{#2}\begingroup\raggedright\endgroup

\end{document}